\documentclass[graybox, nosecnum]{svmult}

\usepackage{mathptmx}       
\usepackage{helvet}         
\usepackage{courier}        
\usepackage{type1cm}        
\usepackage{makeidx}         
\usepackage{graphicx}        
\usepackage{multicol}        
\usepackage[bottom]{footmisc}
\usepackage[flushleft]{threeparttable}
\usepackage{hyperref}        
\usepackage{soul}            
\hypersetup{colorlinks=true,urlcolor=blue}
\usepackage[square,numbers]{natbib}
\usepackage{amssymb}
\usepackage{changes}
\usepackage{aas_macros}

\makeindex             

\begin{document}
\title*{Pulsar Wind Nebulae}
\author{A.M.W. Mitchell and J. Gelfand}
\institute{Alison Mitchell \at Friedrich-Alexander-Universit\"at Erlangen-N\"urnberg, Erlangen Centre for Astroparticle Physics, 
Nikolaus-Fiebiger-Str. 2, D-91058 Erlangen, Germany \\
\email{alison.mw.mitchell@fau.de}
\and Joseph Gelfand \at NYU Abu Dhabi, PO Box 129188, Abu Dhabi, UAE \\
\email{jg168@nyu.edu}}

\maketitle
\abstract{Pulsar Wind Nebulae (PWNe), structures powered by energetic pulsars, are known for their detection across the entire electromagnetic spectrum, with  diverse morphologies and spectral behaviour between these bands. The temporal evolution of the morphology and spectrum of a PWN depends strongly on the properties of the associated neutron star, the relativistic outflow powered by its rotational energy, and surrounding medium, and thereby can vary markedly between objects. Due to the continuous, but decreasing, injection of electrons and positrons into the PWN by the pulsar, the brightness and spectral variation within and amongst their wind nebulae reflect the magnetic field structure and particle transport within the PWN. This can include complex motions such as reverse flows or turbulence due to shock interactions and disruption to the nebula.  During the last stage of the PWN's evolution, when the neutron star moves supersonically with respect to its environment, the escape of accelerated particles into the surrounding medium creates an extensive halo evident in very-high-energy gamma-rays. This chapter describes some of the identifying characteristics and key aspects of pulsar wind nebulae through their several evolutionary stages. }
\section{Keywords} 
Pulsar -- Pulsar Wind Nebula -- acceleration -- magnetic field 


\newpage
\section{Introduction}
Emission from pulsar wind nebulae (PWNe), the structures formed by the interaction between the highly relativistic outflow (``wind'') generated by a neutron star and the surrounding medium, has been detected across the electromagnetic spectrum -- from the lowest frequency $\nu \lesssim 100~{\rm MHz}$ radio waves (e.g., \cite{driessen18}) to extremely high energy $\gamma$-rays (photon energy $E_\gamma \sim 10^{15}~{\rm eV}$; \cite{lhaaso21}).  The detection of such high energy photons from PWNe requires that particles are accelerated to at least PeV energies in these sources, and significant fraction of Galactic TeV sources associated with PWNe (e.g., \cite{hess_pwn_pop, hess18}) suggests these objects are an important source of cosmic ray leptons.  However, the physical mechanisms responsible for producing these high energy particles, and the physical conditions under which this acceleration occurs, remains unknown.  This requires relating the observed properties of a PWN -- e.g., its angular extent and broadband spectral energy distribution (SED) -- to the physical properties of the emitting plasma.  This entails understanding the evolution of a PWN, and the processes responsible for its emission described below.  We first present a physical description of a PWN, and then discuss how the properties of these sources are expect to evolve with time.  We then describe the observed properties of PWNe in the framework of this physical description of their evolution, and then discuss some of the open questions in this field.

\section{Physical Description of a PWN}
\label{sec:phys_model}
\begin{figure}[tb]
    \begin{center}
    \includegraphics[height=0.5\textheight]{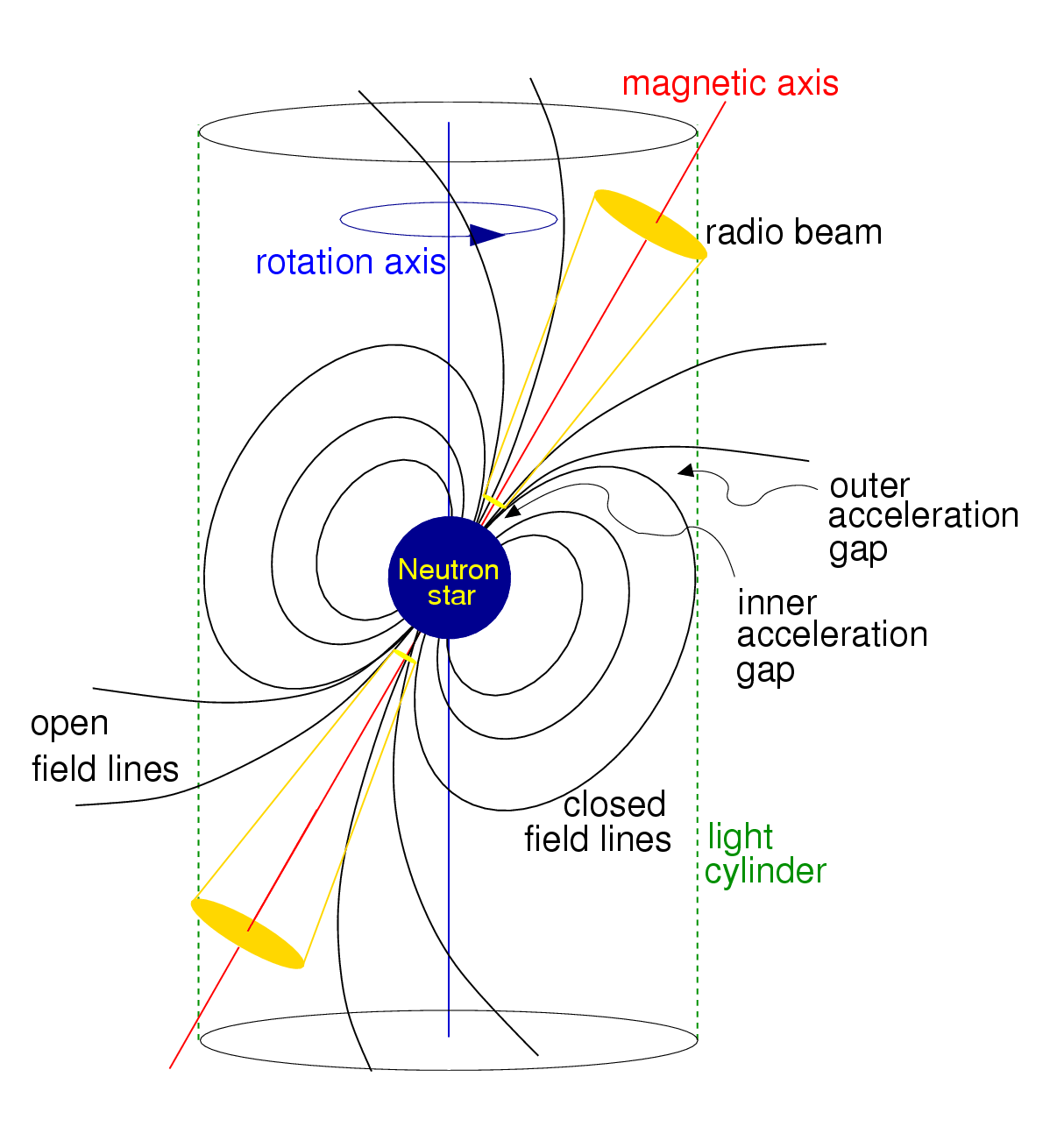}
    \includegraphics[height=0.15\textheight]{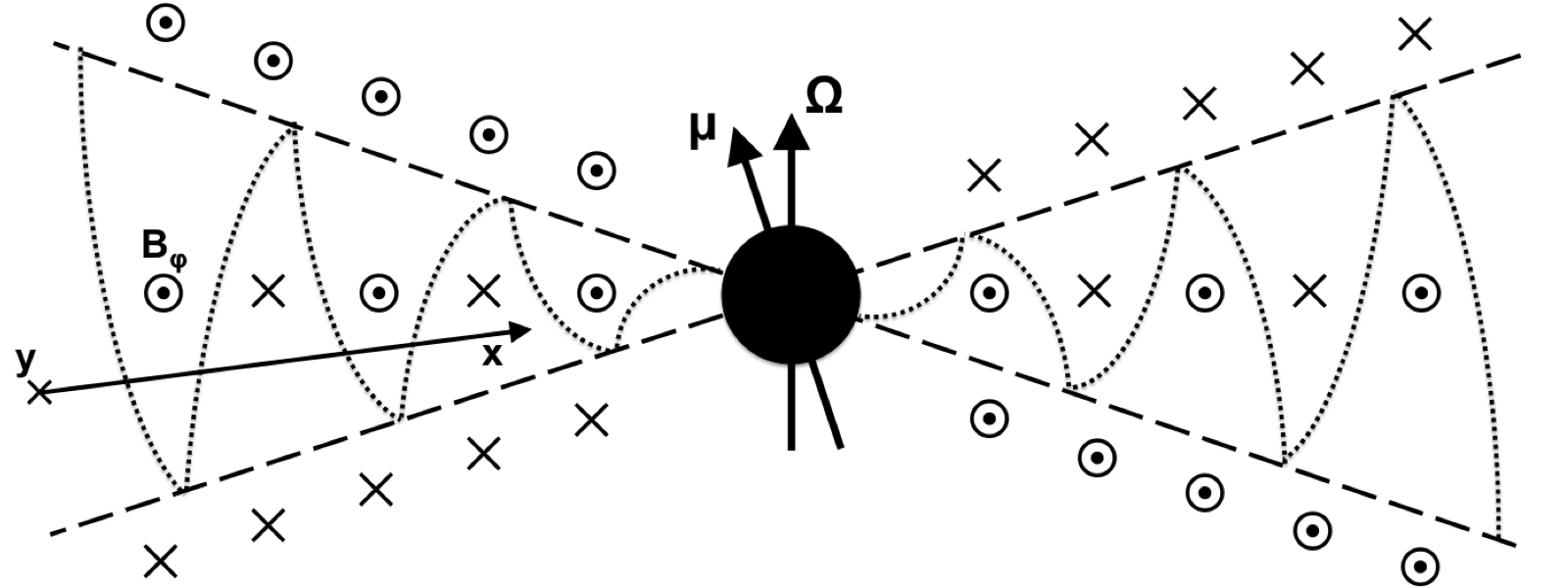}
    \end{center}
    \caption{{\it Top}: Schematic diagram of pulsar magnetosphere, reprinted with permission from \cite{lorimer12}. {\it Bottom}: Schematic diagram of pulsar wind between light cylinder and termination shock.  Reprinted with permission from \cite{mochol17, sironi17}.}
    \label{fig:magnetosphere}
\end{figure}

Even before their discovery, \cite{pacini68} speculated that neutron stars -- the compact object created by the gravitational collapse of a massive star's Fe core whose formation triggers a core-collapse supernova (e.g., \cite{zwicky38} and reference thereafter) -- were responsible for powering the sources now commonly referred to as PWN\footnote{Historically, such sources were often referred to as ``plerions'' \cite{weiler78, weiler78b}, but usage of this term has diminished in recent years as our physical understanding of these objects has improved.}.  These neutron stars can often have extremely strong ($B\gtrsim10^{12}~{\rm G}$) surface magnetic fields, whose field lines are anchored to locations on the neutron star's outer crust.  As a result, near the surface of the neutron star its magnetic field rotates with the same angular frequency $\omega$, and around the same rotation axis $\overrightarrow{\Omega}$, as the neutron star's surface. 
The co-rotation of the neutron star's outer crust and magnetic field near the surface generates an extremely strong electric potential $\Phi$ along the magnetic poles, which creates a powerful current of charged particles (primarily electrons and positrons, $e^{\pm}$, though ions are possibly an energetically important component as well; see \cite{arons94} for a review) who primarily travel along field lines in the magnetosphere (e.g., \cite{goldreich69}; see Figure \ref{fig:magnetosphere}).  

Co-rotation with the neutron star surface at a distance $R$ from the rotation axis $\overrightarrow{\Omega}$ requires a speed $v_{\rm co-rot}=\omega R$. 
Since particles, and magnetic field lines cannot travel faster than the speed of light $c$, such co-rotation is only possible for distances where $v_{\rm co-rot} < c$, or $R < R_{\rm LC}$, where the radius of the light cylinder $R_{\rm LC}$ is:
\begin{eqnarray}
    R_{\rm LC} & = & \frac{c}{\omega}.
\end{eqnarray}
Field lines which extend beyond $R>R_{\rm LC}$ therefore can only be connected to the neutron star surface at one point, and can extend  $R\rightarrow\infty$ far away.  Particles produced along these ``open'' field lines will leave the magnetosphere, and the torque exerted on the neutron star surface, causing its rotational energy $E_{\rm rot}$ to decrease with time.  While the spin-down luminosity $\frac{dE_{\rm rot}}{dt} \equiv \dot{E}_{\rm rot}(t)$ of a pulsar is sensitive to the physical properties of its magnetosphere (e.g., \cite{gruzinov05}), it is typically 
approximated as (e.g., \cite{goldreich69, lorimer12}):
\begin{eqnarray}
    \label{eqn:psr-edot}
    \dot{E}_{\rm rot}(t) & \approx & \dot{E}_{\rm rot,0} \left(1 + \frac{t}{\tau_{\rm sd}}\right)^{-\frac{p+1}{p-1}}
\end{eqnarray}
where  $t$ is the age of the neutron star, $\dot{E}_{\rm rot,0} \equiv \dot{E}_{\rm rot}(t=0)$ is its initial spin-down luminosity, $\tau_{\rm sd}$ is the spin-down timescale for the neutron star, and $p$ is the braking index, defined as $\dot{\omega} \propto \omega^p$.  If the surface magnetic field of a pulsar is a pure dipole whose strength is constant with time (neither of which are true, e.g. \cite{gruzinov05, broderick08}), then $\dot{E}_{\rm rot}$ can be associated with the magnetic dipole radiation of this field in which case:
\begin{itemize}
    \item the braking index $p \equiv 3$,
    \item the measured period $P \equiv \frac{2\pi}{\omega}$ and period-derivative $\dot{P}$ of the neutron star provides an estimate of its surface (dipolar) magnetic field strength $B_{\rm sd}$ (e.g., \cite{lorimer12,condon16} and references therein):
    \begin{eqnarray}
    \label{eqn:bsd}
    B_{\rm sd} & = & 3.2\times10^{19}~\sqrt{P\dot{P}}~{\rm G}
    \end{eqnarray}
    as well as an approximation of its age, assuming an initial spin period $P_0 \ll P$, often referred to as the characteristic age $t_{\rm ch}$ \cite{lorimer12, condon16}:
    \begin{eqnarray}
     \label{eqn:tch}
     t_{\rm ch} & \equiv & \frac{P}{(p-1)\dot{P}} = \frac{P}{2\dot{P}}.
    \end{eqnarray}
\end{itemize}
Under these assumptions, the spin-down timescale $\tau_{\rm sd}$ and age $t$ of the neutron star are related by:
\begin{eqnarray}
 \label{eqn:tausd}
 \tau_{\rm sd} & = & \frac{2t_{\rm ch}}{p-1}-t.
\end{eqnarray}

When the plasma leaves the magnetosphere, it is initially in the form of a strongly magnetized (ratio of the magnetic to total power of this outflow $\eta_{\rm B} \approx 1$ or ratio of particle to magnetic power $\sigma \gg 1$), equatorial wind composed of regions of alternating magnetic polarity, or ``magnetic stripes'' (Figure \ref{fig:magnetosphere}; e.g., \cite{coroniti90, lyubarsky02, mochol17, sironi17} and references therein).  The confinement of ``pulsar wind'' by the surrounding medium, whatever it may be, creates a ``termination shock'' (e.g., \cite{kennel84, slane17}) where this initially highly relativistic (bulk Lorentz factors $\Gamma_0 \sim 10^5$), extremely low pressure outflow is converted to a much lower (bulk) velocity, (relatively) high pressure plasma.  At this termination shock, a significant fraction of the initial magnetic energy of this wind is converted to the particle energy, such that the post-shock plasma is particle dominated ($\eta_{\rm B} \ll 1$, $\sigma \sim 1$; e.g., \cite{kennel84}).  While the physical mechanism responsible for this transformation is currently poorly understood, magnetic reconnection near the termination shock (e.g., \cite{lyubarsky01}) and/or kink instabilities in the post-shock magnetic field (e.g., \cite{porth13}) are believed to significantly contribute to this process.  The dissipation of the magnetic energy of the pulsar wind will also accelerate particles to high energies.  In addition to the ``standard'' Fermi acceleration mechanism thought to occur at strong shocks (e.g. \cite{lu21}), magnetic reconnection at the termination shock is also expected to contribute significantly to the population of accelerated particles (e.g., \cite{sironi11, sironi14}), with the maximum energy of these particles strongly depends on the properties (such as the width of magnetic stripes) and composition of the pre-shock wind (e.g., \cite{arons12, sironi13, lemoine15}).  Measurements of the SED of PWNe suggests that the spectrum of particles accelerated at the termination shock, and injected into the PWN, is often well described by a broken power-law (e.g., \cite{bucciantini11, gelfand15, hattori20, tanaka10, tanaka11, torres13} and references therein and citations thereafter):
\begin{eqnarray}
\label{eqn:bpl}
\frac{d\dot{N}}{dE} & = & \left\{
\begin{array}{cc}
\dot{N}_{\rm break}\left( \frac{E}{E_{\rm break}}\right)^{-p_1} &~~ E_{\rm min} \leq E \leq E_{\rm break}  \\
\dot{N}_{\rm break}\left( \frac{E}{E_{\rm break}}\right)^{-p_2} &~~ E_{\rm break} \leq E \leq E_{\rm max}
\end{array}
\right.
\end{eqnarray}
where typically $p_2$ has a value comparable to that expected from Fermi acceleration at a strong shock ($p_2 \sim 2.5$; .e.g \cite{berezhko99, keshet05, ellison13, ellison16}) while $p_1$ is often considerable smaller, with the harder particle spectrum injected at lower energies often interpreted as evidence for magnetic reconnection.  While observational constraints on $E_{\rm max}$ are typically quite poor, for many PWNe their SED suggests this quantity is of order $\sim{\rm PeV}$ (e.g., \cite{bucciantini11, gelfand15,hattori20,kim20,martin12,slane12,torres13,lhaaso21,2022A&A...660A...8Breuhaus,2022ApJ...930L...2D_uhepwn}).

\begin{figure}[tb]
    \centering
    \includegraphics[width=0.975\textwidth]{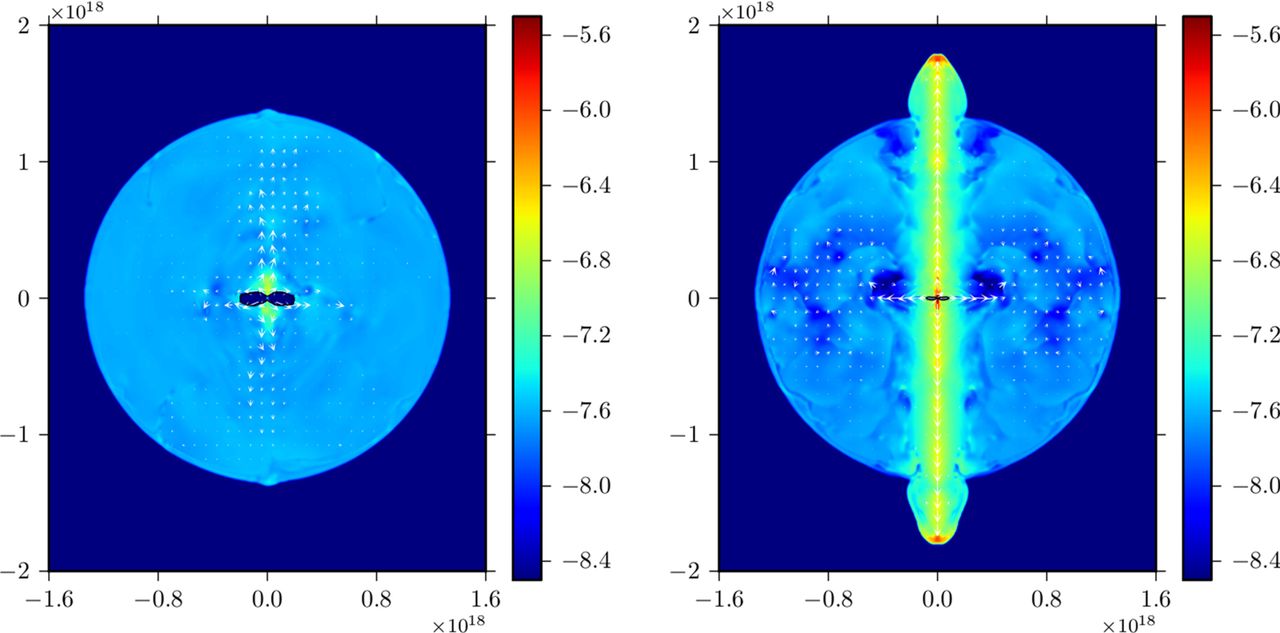}
    \caption{Pressure distribution inside a (Stage 1) PWN as predicted by a 3D MHD simulation of such systems.  Figure reprinted from \cite{porth14}.}
    \label{fig:pwn_pres}
\end{figure}

\begin{figure}[tb]
    \centering
    \includegraphics[width=0.475\textwidth]{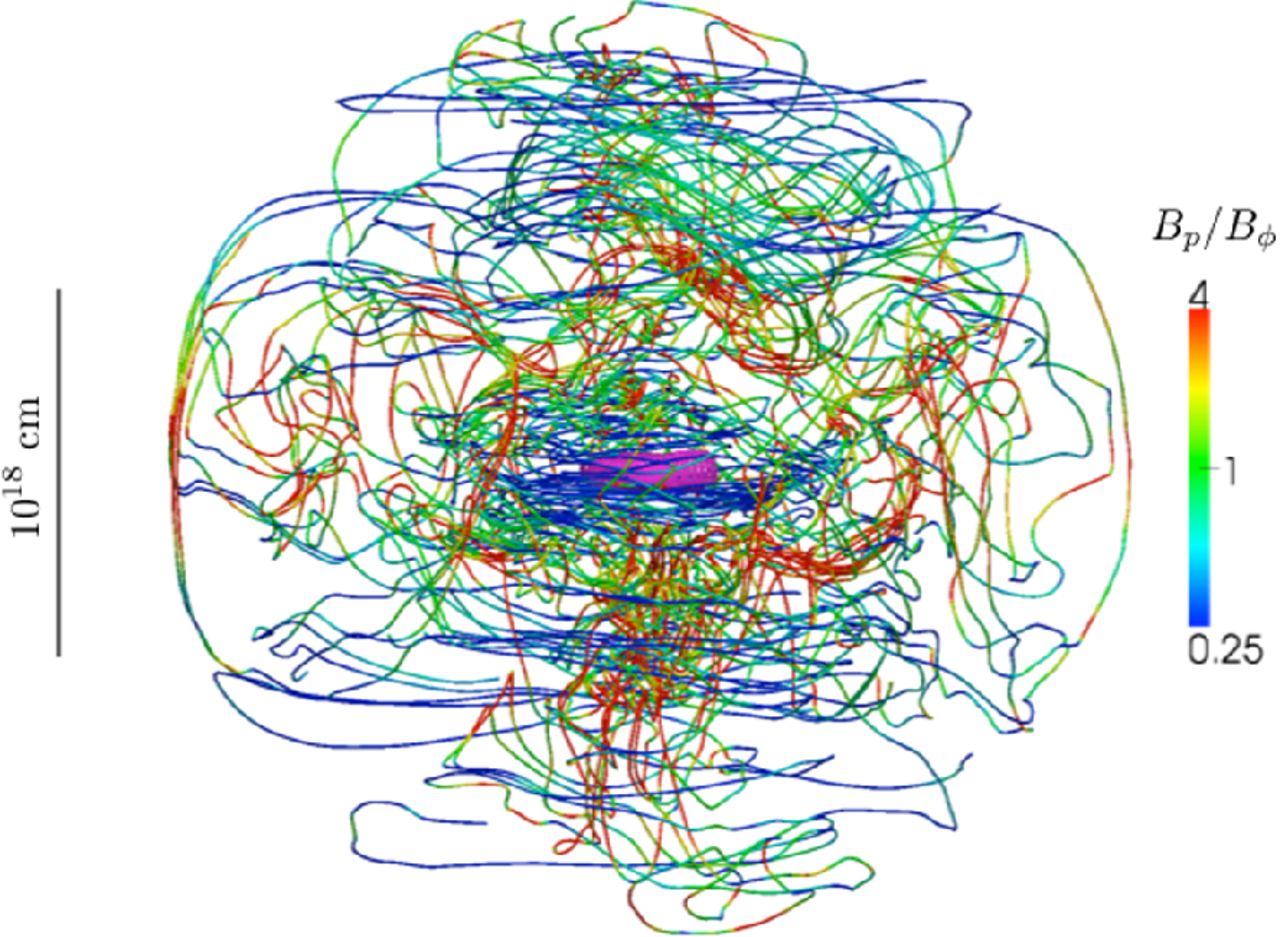}
     \includegraphics[width=0.475\textwidth]{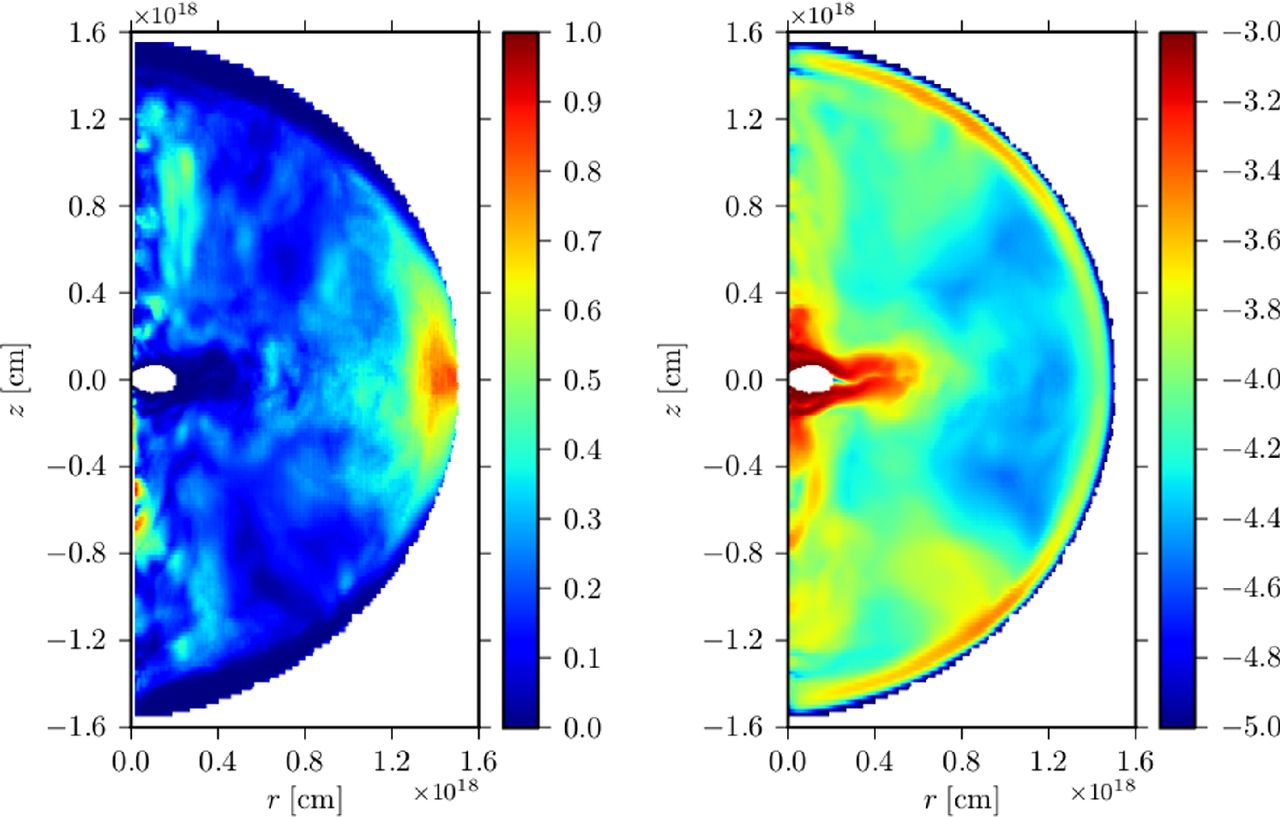}  
    \caption{Magnetic field lines ({\it left}), anisotropy ({\it middle}) and strength ({\it right}) in (Stage 1) PWN as predicted by a 3D MHD simulation of such systems. Figure adapted from \cite{porth14}.}
    \label{fig:pwn_mag}
\end{figure}

The content and emission from a PWN is dominated by this post-shock plasma.  Since it is predominately composed of $e^\pm$ (particle creation or annihilation between the light cylinder and termination shock are thought to be negligible), the dominant emission mechanisms are synchrotron radiation from these leptons interacting with nebula's magnetic field, and the inverse Compton radiation from the interaction between these leptons and lower-energy photons.  These lower energy photons are believed to be dominated by external radiation fields -- e.g., the Cosmic Microwave Background (CMB), thermal emission from local dust, gas, and stars -- with the synchrotron radiation from these leptons (which would result in sychrotron self-compton or SSC emission) an important contributor for only the youngest, more energetic systems (e.g. \cite{torres13}).  Furthermore, the relativistic nature of this plasma means than it has a sound speed $c_s \approx \frac{c}{\sqrt{3}}$ \cite{reynolds84} (for an adiabatic index $\gamma=\frac{4}{3}$) and a comparable fast Alfv\'en speed (e.g., \cite{zrake17}).  As a result, the pressure and magnetic field strength inside a PWN is expected to be roughly uniform (e.g. \cite{reynolds84}), broadly consistent with recent 3D MHD simulations of such objects (Figures \ref{fig:pwn_pres} \& \ref{fig:pwn_mag}; \cite{porth14}).

\begin{figure}
    \centering
    \includegraphics[width=0.98\textwidth]{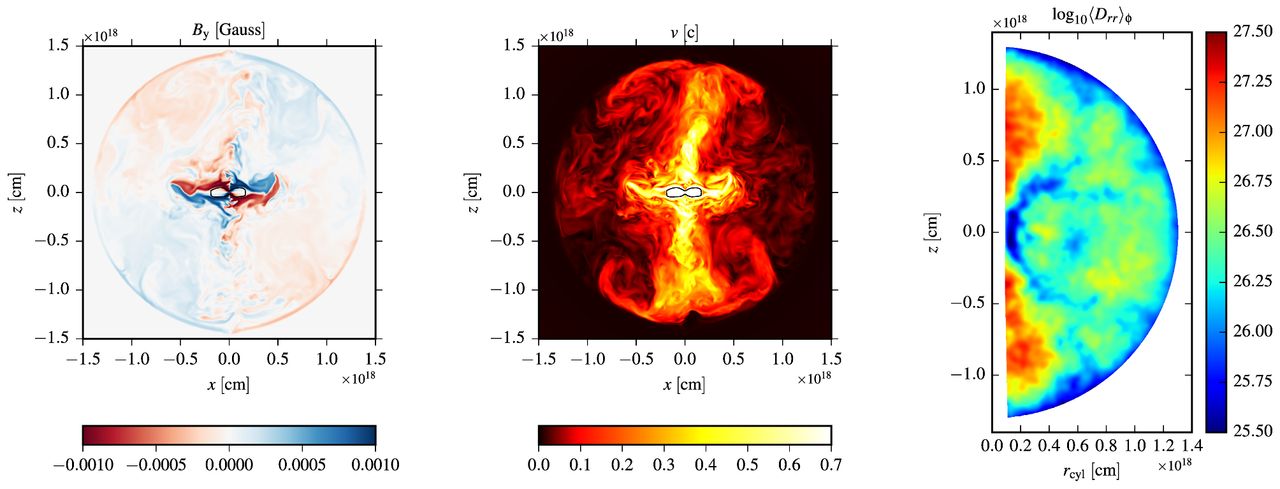}
    \includegraphics[width=0.98\textwidth]{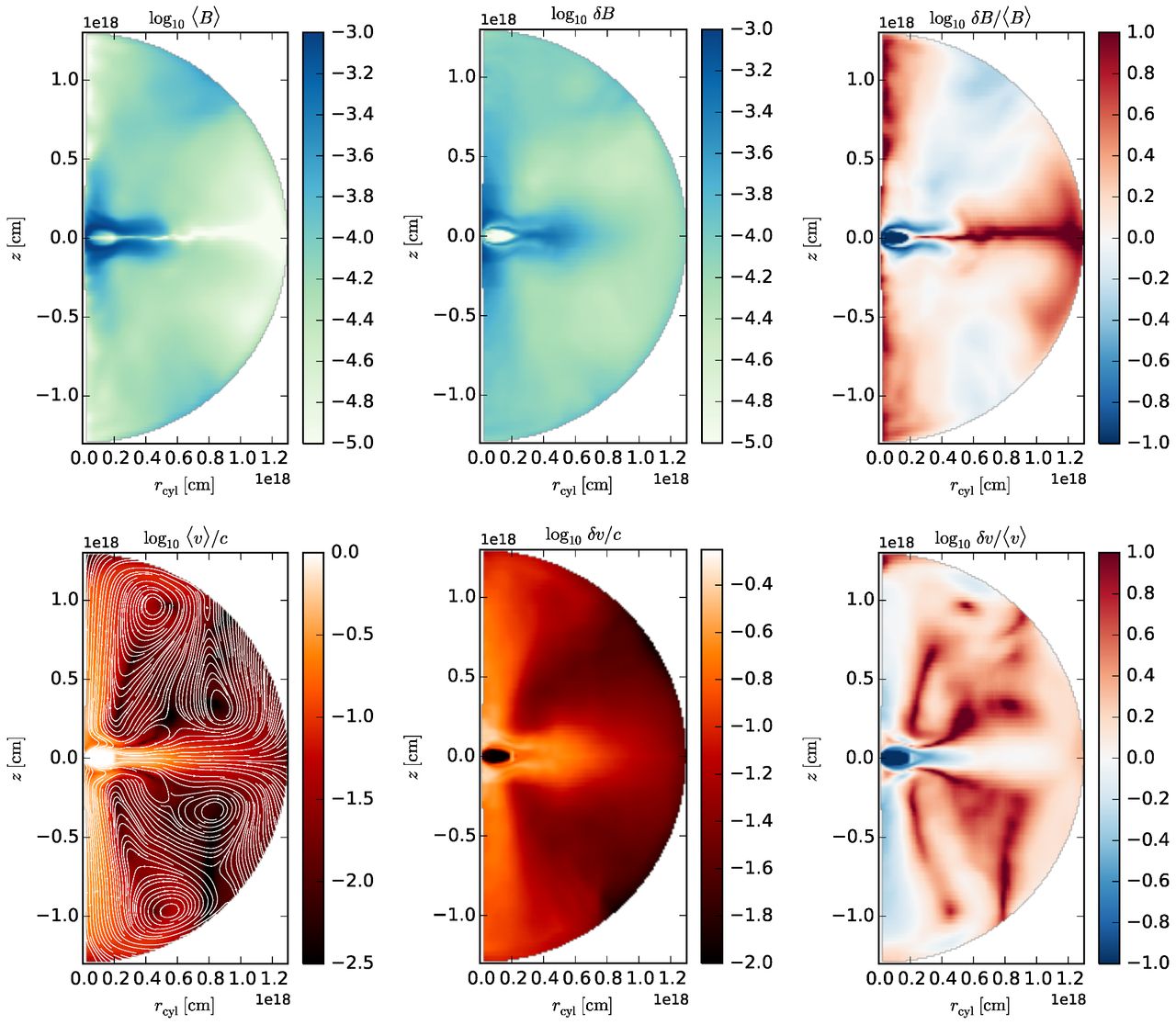}
    \caption{Results from a 3D MHD simulation of a ``Stage 1'' PWN.  The top row shows the strength of the out-of-plane magnetic field ({\it left}), particle speed ({\it middle}), and magnitude of the radial diffusion coefficient ({\it left}).  The second row shows the average magnetic field strength ({\it left}), fluctuation of magnetic field strength ({\it middle}), and relative change in magnetic field ({\it right}) resulting from turbulence inside this simulated PWN.  The bottom row shows the average flux speed ({\it left}), fluctuations in flow speed ({\it middle}), and and relative changes in flow speed ({\it left}) in this simulated PWN.  Note the large scale turbulence motions indicated by the white contours in the bottom-left panel.  All figures reprinted from \cite{porth16}.}
    \label{fig:stage1_turb}
\end{figure}

The observed spatial properties of a PWN -- defined to be the volume of space predominantly comprised of the shocked pulsar wind -- depend on the flow of plasma between the termination shock and the PWN's outer boundary, typically the location of the (``forward'') shock driven into the surrounding medium by the expanding plasma or, in the case of ``bow shock PWNe'', the supersonic motion of the neutron star itself.  The primarily equatorial nature of the pre-shock pulsar wind results in the termination shock having a largely toroidal structure -- believed to be responsible for the ``inner ring'' or torus of X-ray emission observed around some neutron stars (e.g., the Crab Nebula; \cite{weisskopf00}).    As a result, near the termination shock, the post shock flow is primarily toroidal, though instabilities along the edge of the termination shock (often referred to as ``high latitudes'') can redirect material inwards.  The inward material converge along the poles of the termination shock, and the increased pressure at this location results in a high bulk velocity polar outflow commonly referred to as a ``jet'' (Figure \ref{fig:stage1_turb}; e.g.,\cite{lyubarsky02, komissarov03, delzanna04}), though its characteristics are quite different than similar features observed in active galactic nuclei and $\gamma$-ray bursts.

On a microscopic level, as particles leave the termination shock, they rotate around a particular post-shock magnetic field line with a radius $r \sim r_{\rm L}$, the Larmor radius of a particle with energy $E$ in a magnetic field of strength $B$,  As they propagate along a particular field line, a particle can transfer onto another one $\lesssim1r_{\rm L}$ from the original.  As a result, this flow is strongly dependent on the structure of the PWN's magnetic field.  The significant linear polarization of the radio emission detected from a large and diverse set of PWNe (e.g. \cite{schmidt79, kothes06, kothes08, lang10, ng10, ma16, ng17, kothes20, lai22} and references therein) suggests that the magnetic fields in these sources are highly ordered throughout the evolutionary sequence described in the Section above -- though the geometry of this magnetic field can appear significantly different between systems in similar phases.  The recent launch and successful deployment of the {\it Imaging X-ray Polarimetry Explorer} (IXPE; \cite{weisskopf22}) will result in measurement of the magnetic field structure traced by the highest energy particles of the PWN -- important for understanding the underlying acceleration of these particles.  This mission has already produced such measurements for the Crab Nebula (e.g., \cite{bucciantini22}).  Recent models (e.g., \cite{bucciantini18}) and MHD simulations (Figure \ref{fig:pwn_mag}; e.g., \cite{porth14}) are able to qualitatively reproduce such ordered magnetic fields observed inside PWN, though the structure of these simulated magnetic fields are not necessarily consistent with the properties inferred from the polarized radio emission of such systems.

\begin{figure}
    \centering
       \includegraphics[width=0.5\textwidth]{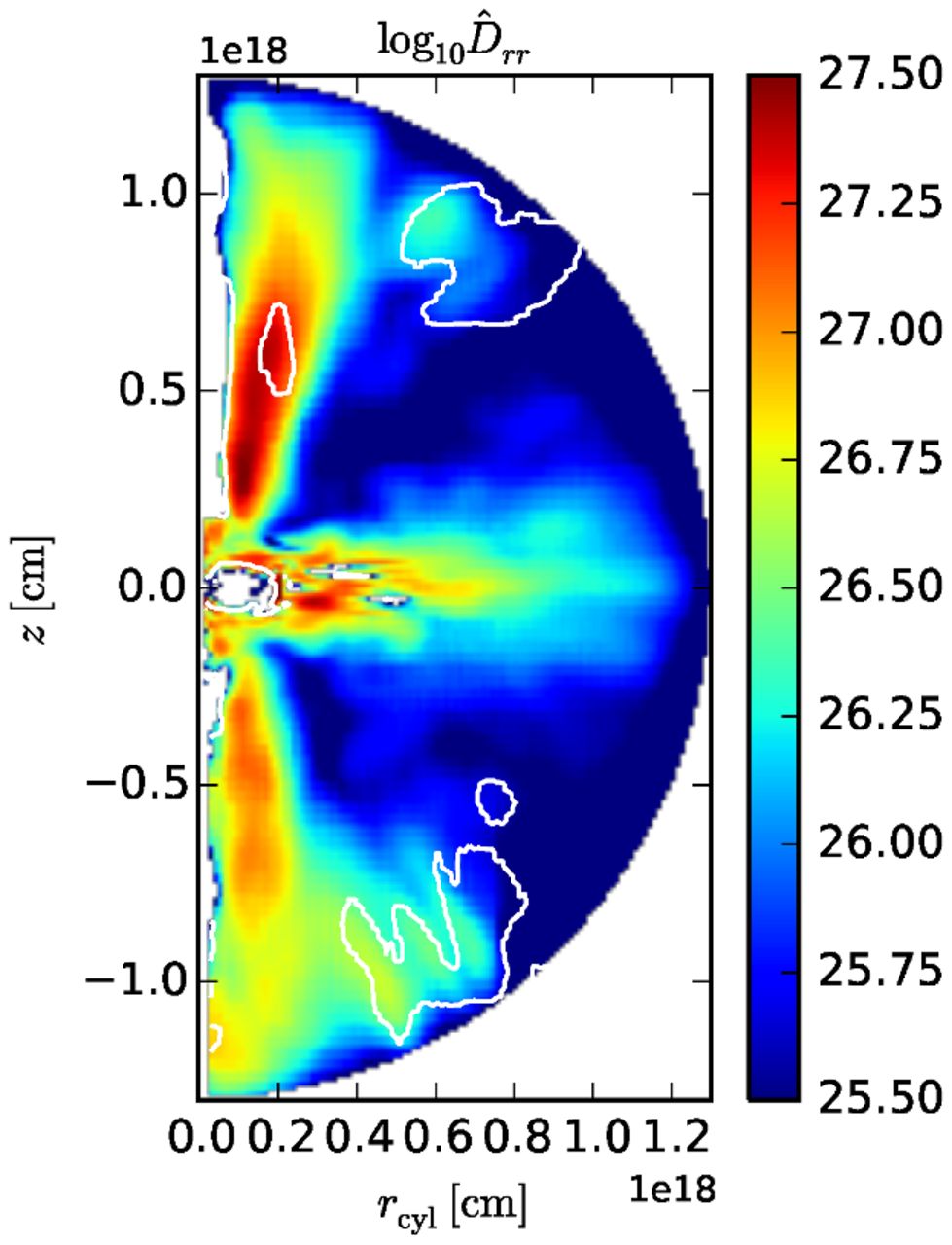}
    \caption{Radial diffusion coefficient associated with the turbulent motions observed in 3D MHD simulation of a PWN. Figure reprinted from \cite{porth16}.}
    \label{fig:stage1_dif}
\end{figure}

On a macroscopic level, this flow can be treated as a combination of advection and diffusion (e.g., \cite{tang12, vorster13}), with a degree of turbulence (e.g., \cite{porth16, zrake17}) resulting from instabilities within the nebular magnetic field (e.g., \cite{begelman98, mizuno11}).  Spatial variations in the strength and geometry of the PWN's magnetic will likely lead to spatial variations in the dominant transport mechanism as well as in the diffusion coefficient (e.g. Figure \ref{fig:stage1_turb} \& \ref{fig:stage1_dif}; \cite{porth16}), which can be used to explain observed variations in the surface brightness and spectrum of PWN (e.g., \cite{tang12, ishizaki17, ishizaki18, hu22}).  Furthermore, the spectral evolution of the particles inside PWN -- and their resultant emission -- is strongly affected by the radiative losses of particles as the travel within the PWN, the adiabatic processes governing the overall interaction of the PWN with its surroundings, and the escape of particles from the PWN into the surrounding medium (e.g., \cite{reynolds84, gelfand09, bucciantini11, martin12, tanaka11, zhu18, zhu21}).  The escape of particles from the PWN is especially important for understanding the diffuse $\gamma$-ray emission observed around an increasing number of neutron stars.  The rate and mechanism by which particles leave the PWN despite strongly on the stage of its evolution, as discussed in the next section.

\section{PWN Evolution}
\label{sec:evolution}

While the physical model of a PWN described above is expected apply throughout the lifetime of the neutron star, their physical characteristics will change significantly during this time.  This evolution can be broadly attributed to two factors:
\begin{itemize}
    \item the decrease in energy deposited into the PWN by the central neutron star with time, as indicated in Equation \ref{eqn:psr-edot},
    \item and, more importantly, changes in the environment of the neutron star and its PWN.
\end{itemize}
Below, we describe the expected changes in the environment of a neutron star, and how these changes impact the properties of the resultant PWN.

\begin{figure}[tb]
    \centering
    \includegraphics[width=0.975\textwidth]{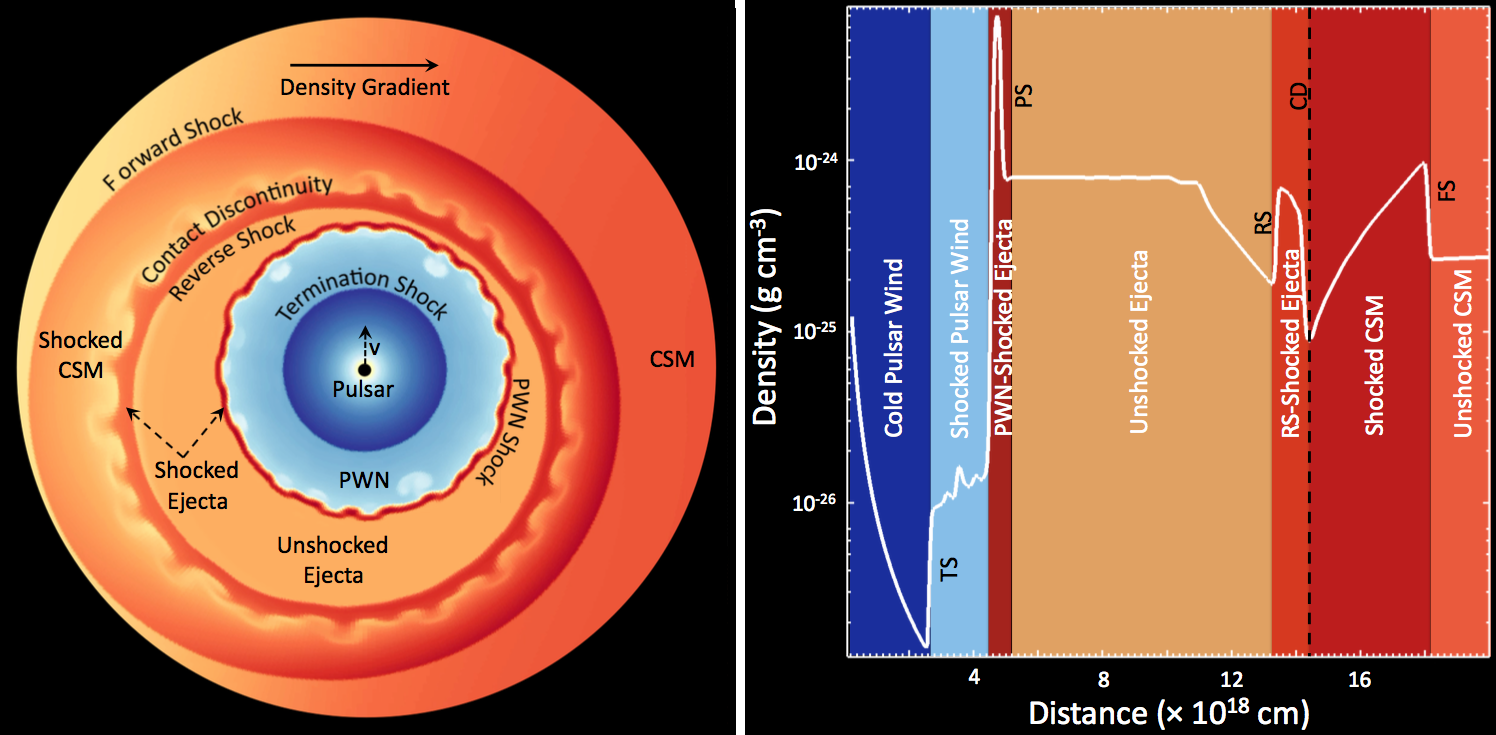}
    \caption{Schematic diagram of the structure of a PWN inside a SNR.  This is the equivalent of Stage 1 in a later figure.  Slane Handbook of Supernovae, ISBN 978-3-319-21845-8. Springer International Publishing AG, 2017, p. 2159
    }
    \label{fig:schematic}
\end{figure}

\begin{figure}[tb]
    \centering
    \includegraphics[height=0.175\textheight]{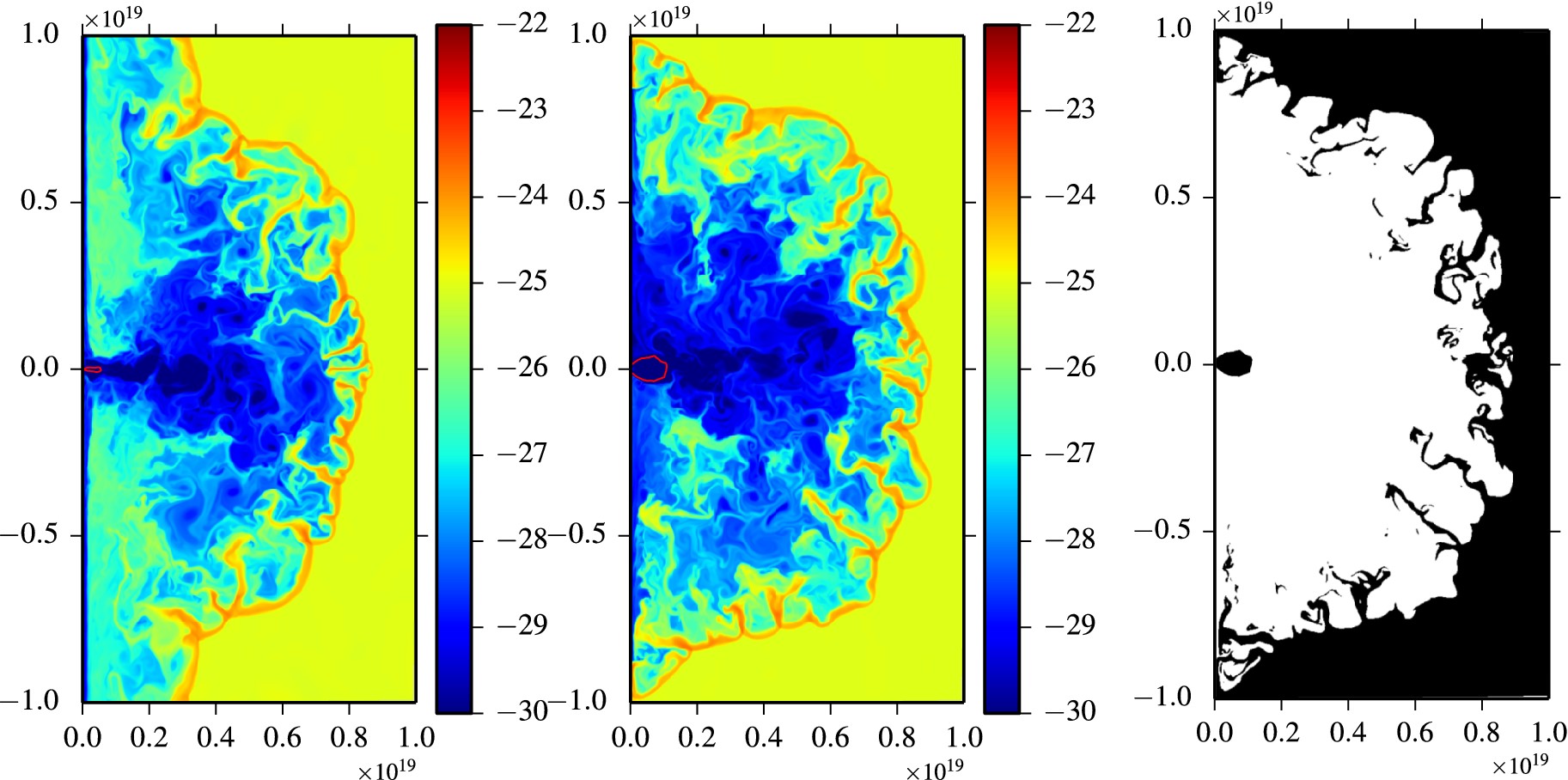}
    \includegraphics[height=0.175\textheight]{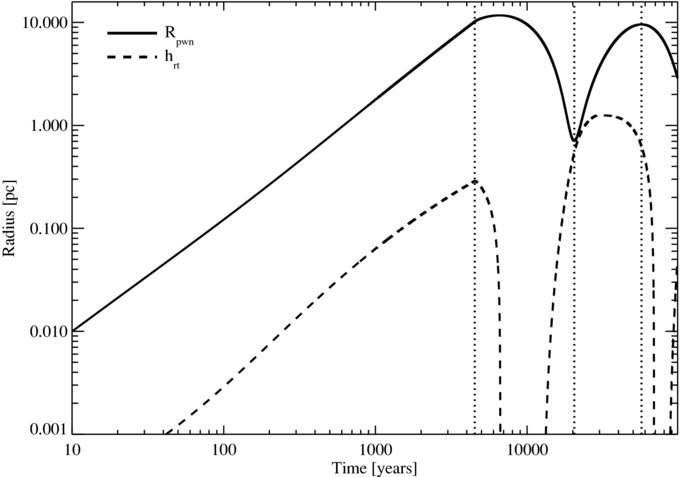}
    \caption{{\it Left}: Simulation of the growth of a Stage 1 PWN with different values of the wind magnetization.  As shown in this figure, the PWN is more elongated in higher magnetization wind, and in both Rayleigh-Taylor instabilities are observed to grow at the interface between the PWN and surrounding SNR.  Figure reprinted from \cite{porth14b}. {\it Right}: Radius $R_{\rm pwn}$ and penetration depth of Rayleigh-Taylor instabilities $h_{\rm RT}$ in a PWN as calculated using the evolutionary model of \cite{gelfand09}.  The calculation of $h_{\rm RT}$ in this model neglects the possible suppression of such instabilities by the nebula's magnetic field.  Figure reproduced by permission of the AAS from \cite{gelfand09}.}
    \label{fig:rt_sim}
\end{figure}

Neutron stars have long been associated with core-collapse supernovae (e.g., \cite{zwicky38, zwicky39}), and are believed to be the remnant of the progenitor star's Fe core whose  gravitational collapse triggered this explosion.  Therefore, initially the neutron star is located in the ``center'' of the supernova remnant (SNR) formed by the expansion of the supernova ejecta into its surroundings (Figure \ref{fig:schematic}; \cite{gelfand09} and many others).  As a result, the PWN is initially surrounded by the slowest moving material produced in this explosion, and low density $\rho$, high pressure $P$ plasma produced at the termination shock is expanding into the high density, essentially pressure-less $P\approx 0$ unshocked ejecta.  The higher pressure inside the PWN causes it to expand within the SNR, sweeping up the surrounding material into a thin shell that confines the pulsar wind inside the PWN.  The shock generated by the PWN's expansion, as well as its emission, can heat dust created inside the SNR, resulting in infrared (e.g., \cite{temim12, kim13, temim17, chawner19, temim19}) and (primarily near-infrared and optical) spectral line (e.g., \cite{sankrit98, zajczyk12, lee19}) emission.  While this shell is expected to be very effective at confining the shocked pulsar wind inside the PWN, the extreme discrepancy in density between the pulsar wind and surrounding ejecta results in the formation of growth of Rayleigh-Taylor instabilities \cite{chandrasekhar61} much may assist in the escape of some particles from this source.  The presence of such instabilities have been used to explain the observed filamentary structure observed around some PWN (e.g., \cite{hester96}), and their development is believed to be sensitive to the rate the PWN expands within the SNR (e.g. \cite{gelfand09, porth14b}) and strength and orientation of the PWN's magnetic field (Figure \ref{fig:rt_sim}; e.g. \cite{bucciantini04, porth14b}) -- with a stronger tangential component to the nebular magnetic field believed to decrease the penetration of these instabilities into the PWN, and inhibit particle escape.

\begin{figure}[tb]
    \centering
    \includegraphics[height=0.2\textheight]{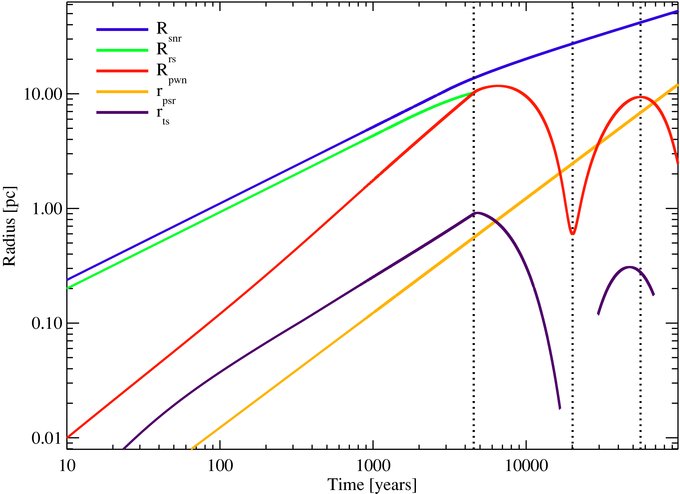}
    \includegraphics[height=0.2\textheight]{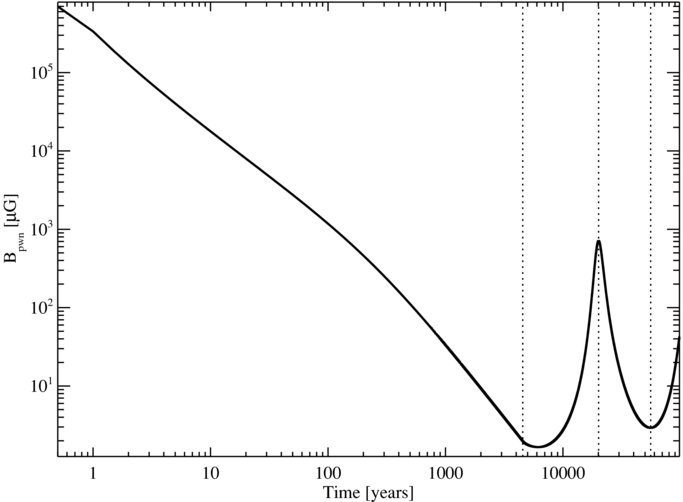}
    \caption{{\it Left}: Radius of the SNR, SNR reverse shock, PWN, termination shock, and location of the central neutron star with time for a PWN inside a SNR. {\it Right}: Strength of the PWN's magnetic field with time.  Both figures reproduced by permission of the AAS from \cite{gelfand09}.}
    \label{fig:pwn_in_snr}
\end{figure}

\begin{figure}[tb]
    \centering
    \includegraphics[width=0.95\textwidth]{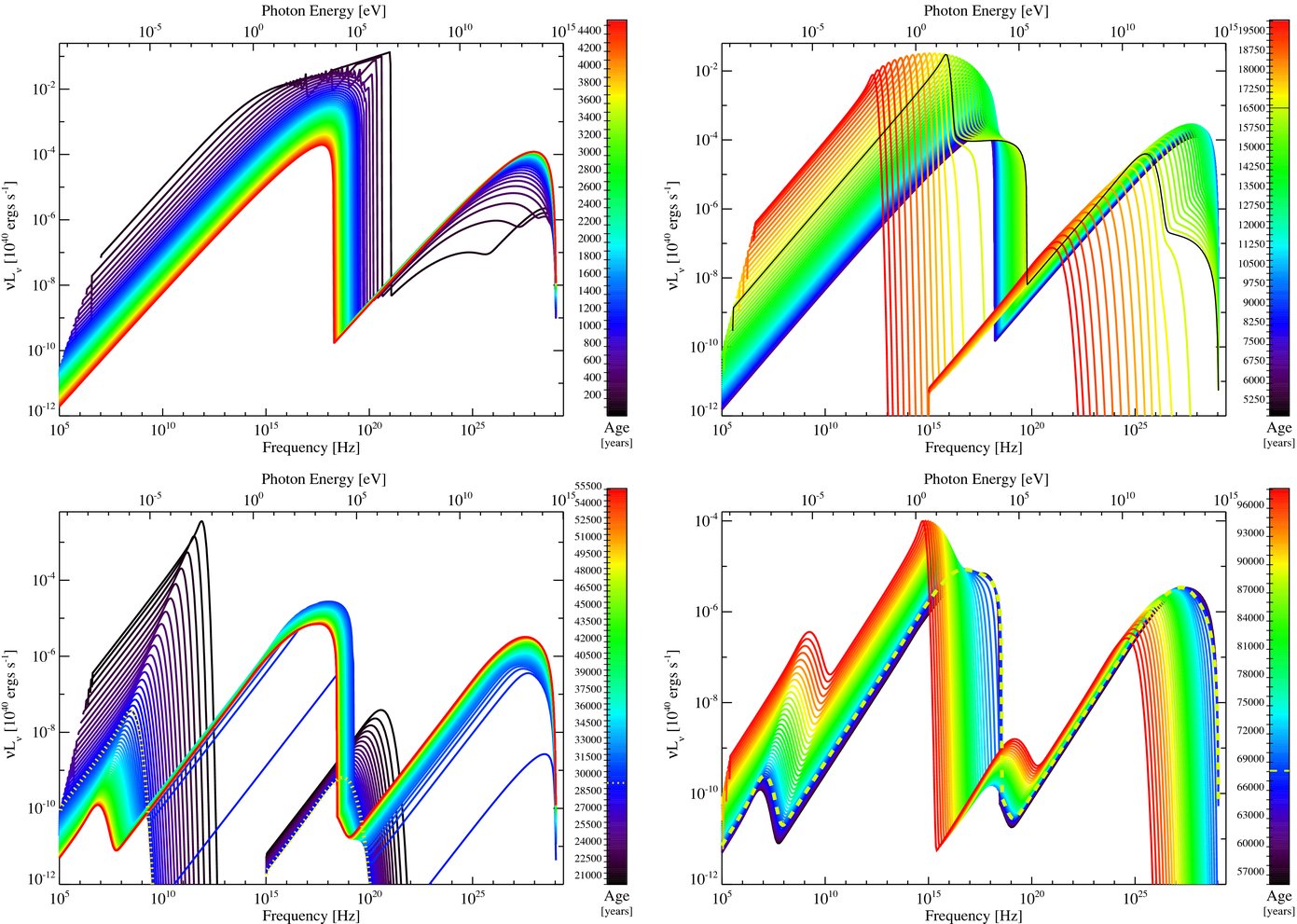}
    \caption{Theoretical evolution of the SED of a PWN inside a SNR for the four phases indicated by the vertical lines in the left panel of Figure \ref{fig:pwn_in_snr}. The upper-left panel displays the SED during the initial free-expansion of the PWN into the SNR, the upper right after the collision with the reverse shock until the end of the first compression, the lower left during the first re-expansion, and then the lower right during the second compression.  Figure adapted from \cite{gelfand09} by permission of AAS.}
    \label{fig:pwn_in_snr_sed}
\end{figure}

During this phase, the radius of the PWN is expected to expand rapidly, increasing as $R_{\rm pwn} \propto t^{\lesssim1.2}$ (Figure \ref{fig:pwn_in_snr}; e.g., \cite{reynolds84, chevalier05, gelfand09}), which results in a significant decrease in the strength its magnetic field strength (Figure \ref{fig:pwn_in_snr}).  The changing physical conditions drive the evolution of the PWN's SED during this initial, free-expansion phase (e.g., \cite{reynolds84, 
gelfand09, tanaka10, bucciantini11}), with the decreasing magnetic field resulting in a decline in the synchrotron emission from the nebula, while the continuing injection of particles and energy by the central pulsar -- which exceeds both the radiative and adiabatic losses of previously injected particles -- into the PWN causes the inverse Compton emission from the PWN to increase with time (Figure \ref{fig:pwn_in_snr_sed}).  The properties of the PWN during this stage are sensitive to the birth properties of the central neutron star and the characteristics of the progenitor supernova, and therefore observations of such PWN have been used to constrain these quantities (e.g., \cite{hattori20})

\begin{figure}
    \centering
    \includegraphics[width=0.975\textwidth]{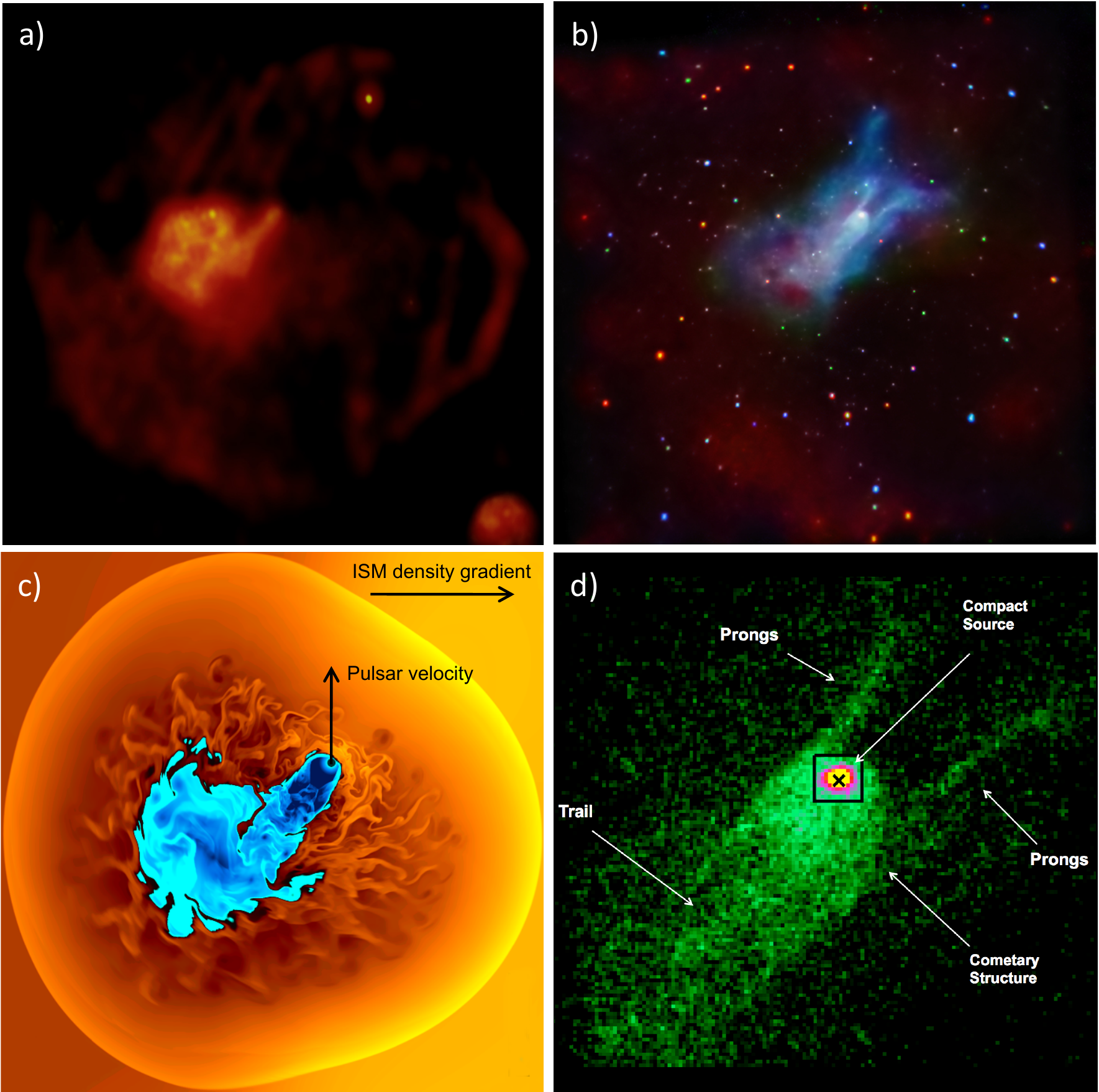}
    \caption{``The Snail'' -- an example of a ``Stage 2'' system where the PWN is in the process of being disrupted by the SNR reverse shock. Slane Handbook of Supernovae, ISBN 978-3-319-21845-8. Springer International Publishing AG, 2017, p. 2159}
    \label{fig:snail}
\end{figure}

The initial expansion of the PWN ends when it collides with the ``reverse shock" driven into the supernova ejecta but interstellar medium shocked and heated at SNR forward shock (see \cite{bandiera21} for a recent calculation for its propagation inside a SNR).  When this occurs, the pressure inside the PWN is much less than its surroundings, resulting in a inwards force on the shell of material surrounding this PWN.  This force first decelerates, and then eventually compresses, the PWN (Figure \ref{fig:pwn_in_snr}; e.g., \cite{reynolds84, vds01, blondin01, bucciantini03,  gelfand09, vorster13}) significantly impacting both the SED (Figure \ref{fig:pwn_in_snr_sed}) and morphology (Figure \ref{fig:snail}) and SED of the PWN.

While the exact impact of the reverse shock of a PWN depends on the particular properties of the system, the following generalities are believed to hold:
\begin{itemize}
    \item This collision is likely to be asymmetric due to the space velocity of the pulsar, which causes it move from the SN explosion site, and gradients in the surrounding medium.  This asymmetry will likely displace the bulk of the PWN from the location of the pulsar (e.g., Figure \ref{fig:snail}; \cite{temim13, temim15, kolb17}), and can possible physically separate the two (e.g., \cite{gelfand09}).  In this case, there would technically be two PWNe inside the SNR -- a ``relic'' composed of particles injected into the PWN before this collision, and a PWN powered by particles recently injected by the neutron star. 
    \item The compression of the PWN will increase the strength of the nebula's magnetic field (Figure \ref{fig:pwn_in_snr}; e.g., \cite{bucciantini03, chevalier05,gelfand09}).  This significantly increases the synchrotron emission from the PWN, to the point that its luminosity can exceed the spin-down luminosity of the neutron star \cite{torres18, bandiera20}.  This increased luminosity results from the decreased synchrotron cooling time of particles within the nebula.  As a result, particles injected into the PWN before it collides with the reverse shock will lose the bulk of their energy during this compression, creating a ``relic population" of low energy particles inside the PWN even if the PWN remains attached to the neutron star.  However, the radiative losses of particles injected into the PWN during this compression will be much lower, and therefore such particles will dominate the observed high energy emission.  This resultant dichotomy in the particle spectrum both introduces spectral ``breaks'' into the SED of the PWN (top right panel of Figure \ref{fig:pwn_in_snr_sed}) and differences in the radio and X-ray morphology of the PWN (e.g., Figure \ref{fig:snail}), since the lower energy particles responsible for the radio emission will reflect the overall extent of the PWN while the more recently injected higher energy particles will be concentrated near the neutron star.
\end{itemize}
The adiabatic compression of the PWN will cause its pressure to increase, until at some point it will exceed that of the surrounding material.  At this point, the low density plasma inside the PWN will again accelerate its much higher density surroundings.  In hydrodynamic models of such systems (e.g., \cite{blondin01, gelfand09}), the rapid growth of Rayleigh-Taylor instabilities expected under these conditions will disrupt the PWN -- injecting all of the particles into the surrounding SNR -- though it unclear if the strong magnetic field inside the PWN when this re-expansion begins (Figure \ref{fig:pwn_in_snr}; \cite{gelfand09}) prevents this disruption from occuring (e.g. \cite{bucciantini04}).

\begin{figure}[tbh]
    \centering
    \includegraphics[width=0.975\textwidth]{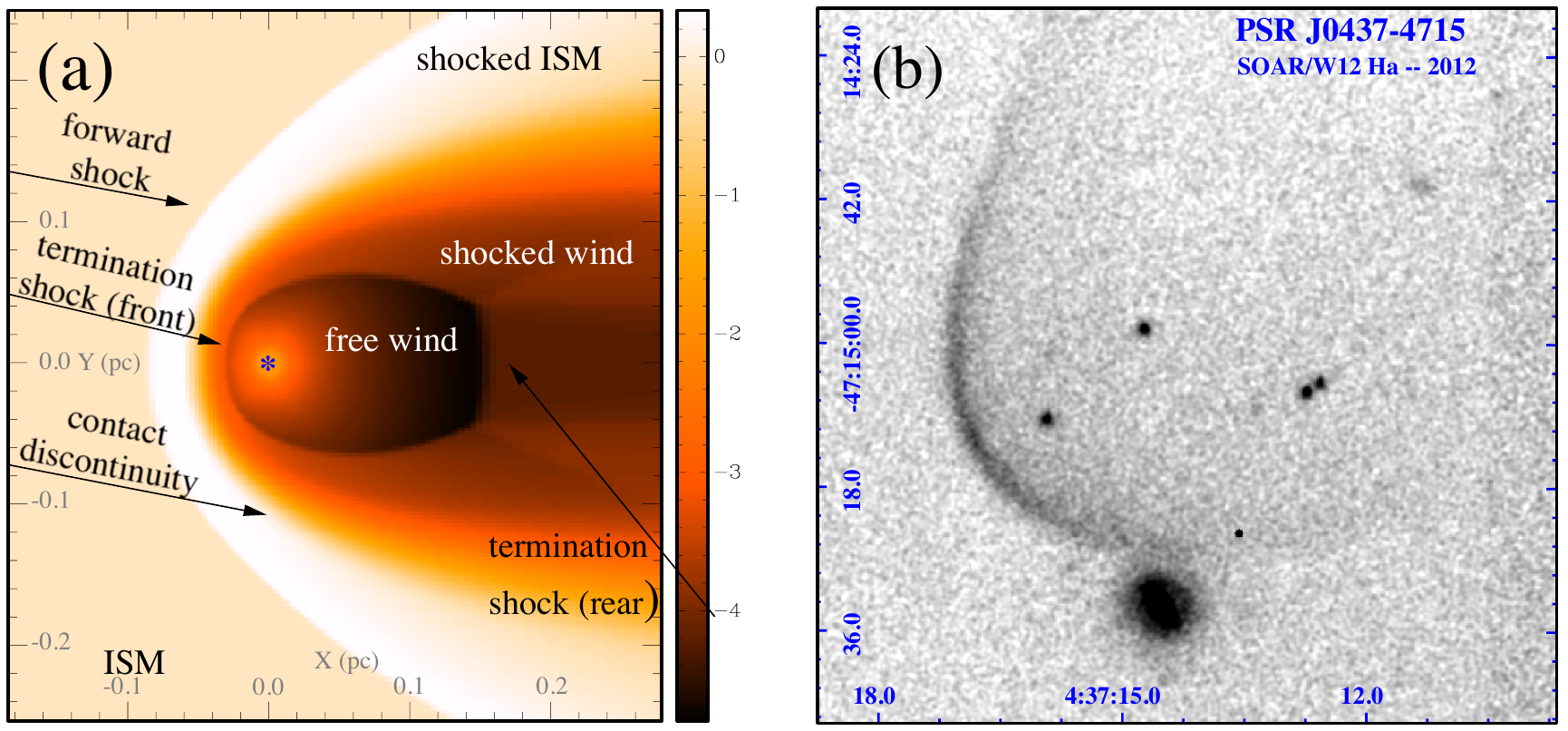}
    \caption{Schematic diagram of a ``Stage 3'' . Slane Handbook of Supernovae, ISBN 978-3-319-21845-8. Springer International Publishing AG, 2017, p. 2159}
    \label{fig:bowshock}
\end{figure}

As the pulsar nears the outer boundary (e.g., \cite{vds04}), and eventually exits, the SNR, it will be moving supersonically relative to its environment, and will develop a bow-shock morphology due to the confinement of the pulsar wind by the ram pressure resulting from this motion.  By this time, the age of the neutron star $t$ is expected to be much longer than its spin-down timescale, in which case the rate its injects energy into its surroundings $\dot{E}_{\rm psr}$ is roughly constant with time.  During this phase, high-energy particles accelerated at the termination shock are expected to propogate into the surrounding ISM through a variety of mechanisms mediated by the interaction between the PWN's and interstellar magnetic fields  (e.g., \cite{bucciantini18b, barkov19, olmi19}).

\begin{figure}
    \centering
    \includegraphics[width=\textwidth]{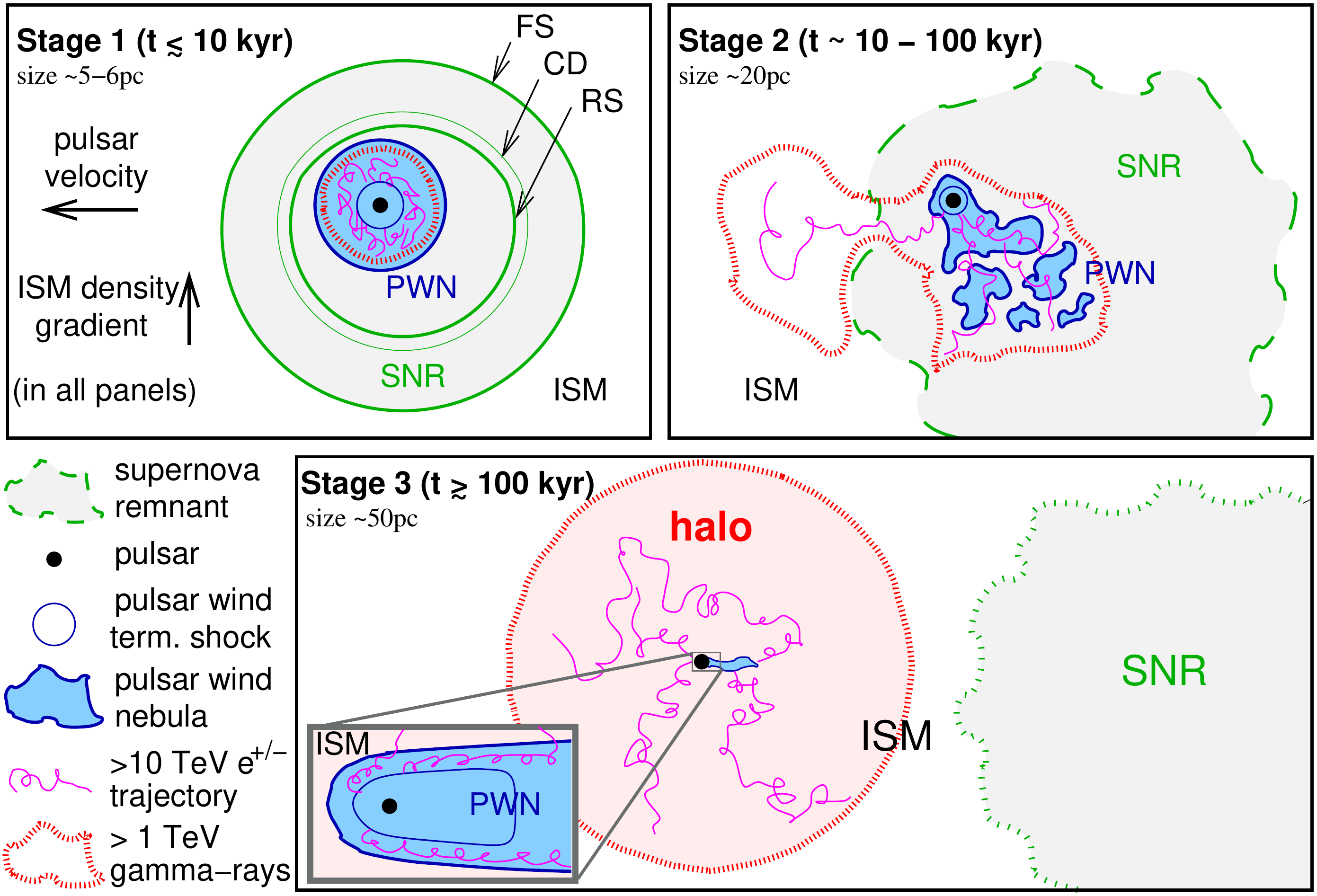}
    \caption{Schematic illustrating the evolutionary stages of a PWN environment. The pulsar velocity is towards the left and ISM density gradient upwards in all three panels. The region producing high energy ($>1$\,TeV) gamma-rays through inverse Compton scattering is determined by the trajectories of the correspondingly energetic ($>10$\,TeV) electrons and positrons. With increasing age, the physical size of the gamma-ray emitting region also increases as indicated. X-ray emission is expected from the region bounded by the canonical pulsar wind nebula, shown in blue. 
    Credit: Giacinti et al. A\&A, 636, A113, (2020), reproduced with permission © ESO.
    }
    \label{fig:pwn_sketch}%
\end{figure}

In summary, the evolution of a PWN can be described as consisting of three stages (Figure \ref{fig:pwn_sketch}):
\begin{enumerate}
    \item {\bf Stage 1} -- an initial expansion into the cold, slow moving ejecta in the center of the SNR, where the shocked pulsar wind is confined by a shell of swept-up material.
    \item {\bf Stage 2} -- which begins when the PWN begins to interaction with the material inside the SNR heated by the reverse shock, which initially compresses, and then possibly disrupts the PWN, at which point a significant fraction of the high-energy particles accelerated with the PWN are injected into the surrounding SNR.
    \item {\bf Stage 3} -- which begins when the neutron star begins to move supersonically with respect to its surroundings.  High-energy particles within the resultant bow shock PWN are expected to escape into the surrounding medium.
\end{enumerate}
As described above, in theses latter stages, a higher fraction of the total high-energy particles accelerated inside a PWN will be found outside this structure, and diffuse within the ISM.  Therefore, their emission will extend well beyond the hydrodynamic boundaries of the PWN.  This emission is likely to be most prominent in $\gamma$-rays, since the background photons needed to produce inverse Compton radiation are abundant while the weak interstellar magnetic fields will likely result in the synchrotron radiation having a surface brightness too low to be observed by current facilities.

\section{Observational Signatures and Notable PWNe}


The dominant emission mechanisms of synchrotron radiation and inverse Compton (IC) scattering lead to MWL SEDs that can be fairly well understood as a function of age of the responsible particle population.  
At X-ray energies, the synchrotron radiation is generated from young, high energy particles recently released into the pulsar wind. The X-ray morphology traces the magnetic field structure around the pulsar, typically in tails or toroidal forms. At gamma-ray energies, the IC radiation can be generated by much older particles, with lower energy yet that continue to up-scatter photons from ambient radiation fields. 

Especially at gamma-ray energies, it is often challenging to distinguish the supernova remnant and plerionic components of a source. Composite supernova remnants ``spp'' for which this distinction has so far not been possible therefore form a sub-category of Galactic gamma-ray sources \cite{2018A&A...612A...1HGPS,2020ApJS..247...33A_4fgl}. 

Figure \ref{fig:pwn_sketch} provides a sketch of PWN evolution, with a focus on the TeV emission, following well-understood stages yet incorporating the more recently established halo phenomenon \cite{2006ARA&A..44...17Gaensler_Slane,slane17,2020A&A...636A.113G}. 
Although approximate age ranges are given, we note that the transition between evolutionary stages depends on the properties of the SN itself and of the surrounding ISM. 

Where density gradients are present in the surrounding ISM and/or the pulsar has a significant kick-velocity, the PWN-SNR system can become highly asymmetric with arbitrary morphology. 

At late evolutionary stages (e.g. third panel of figure \ref{fig:pwn_sketch}), the remaining PWN is comprised only of the most recently accelerated energetic particles. The continuous acceleration of electrons and positrons maintains a small-scale PWN over long times, even after the pulsar has escaped the parent SNR. However, a much larger relic halo of escaped electrons and positrons forms around the pulsar in the ISM. Inverse Compton scattering of these leptons on ambient radiation fields produces a large-scale gamma-ray halo $\sim$10-100\,pc in size. 
Extended, diffuse X-ray emission corresponding to mid-evolution PWNe (i.e. the second panel of figure \ref{fig:pwn_sketch}) has been seen in several systems \cite{2009PASJ...61S.189U_SuzakuJ1825,2010ApJ...719L.116B_xray_pwne}.
Typically, the gamma-ray halo for late evolutionary stages extends out to sizes far in excess of the remaining PWN. 
Whether or not pulsar halos are also accompanied by large-scale diffuse X-ray emission remains to be investigated and verified. 

Although a strict classification based on the characteristic age of the pulsar is tempting, it is worth bearing in mind that the age is an inherently uncertain quantity. The age is determined from the measured pulsar period and period derivative from equation (\ref{eqn:tch}), 
where the characteristic age $\tau_c$ is obtained by crudely assuming that the braking index takes on a value of $3$, such that $\tau_c=\frac{P}{2\dot{P}}$. In fact, this assumption is only valid for a pure magnetic dipole radiation; a value of the braking index can be measured via $n=\frac{f\ddot{f}}{\dot{f}^2}$. For pulsars with a measured value of the braking index, this is typically $<3$ indicating that the magnetic field structure is not a perfect dipole \citep{2015MNRAS.446..857Lyne}. 

Additionally, the transition between evolutionary stages is not determined solely by the age of the system, but is highly influenced by the properties of the SN, the density of the surrounding medium, and in particular the time of the collision between the Reverse Shock and the PWN. Similarly, the time at which the pulsar escapes the SNR (forming a bow shock nebula) depends on the initial explosion and kick velocity, as well as the SNR evolution in the surround medium. Large uncertainties are also inherent in the spin-down timescale $\tau_{\rm sd}$, which can also vary by orders of magnitude, up to $10^4$ years (e.g., \cite{hattori20}).
Therefore, $\tau_c$ should generally be used with caution as a means to classify PWN evolutionary stages, as large uncertainties may be inherent in the obtained $\tau_c$ and its relevance to the current state of the PWN evolution. 

\subsection{Radio}
As shown in Figure \ref{fig:pwn_in_snr_sed}, the radio emission from a PWN is primarily produced by synchrotron radiation from -- for the $B_{\rm pwn}\sim\mathcal{O}(\mu {\rm G})$ magnetic field strengths expected inside these objects (e.g., Figure \ref{fig:pwn_in_snr}) -- electrons and positrons with GeV energies.  Due to their long radiative lifetime (which can exceed the current age of the Universe; e.g., \cite{hattori20}), these particles are expected to fill the volume of the PWN.  The fairly uniform particle distribution (as indicated by the nearly constant particle pressure expected inside these sources; Figure \ref{fig:pwn_pres}) and magnetic field strength (Figures \ref{fig:pwn_mag} \& \ref{fig:stage1_turb}) suggests that synchrotron emissivity of a PWN should also be fairly uniform inside such sources.  The low lepton densities inside a PWN result in a low optical depth due to free-free absorption, and therefore are optically thin at radio wavelengths.  These two factors combined suggest a PWN should have a centrally-filled morphology at radio wavelengths, which is indeed observed and the basis of the ``plerionic'' designation sometimes used for such objects (\cite{weiler78, weiler78b}).  However, since the synchrotron emission from a particle is intrinsically beamed along its direction of motion, in the case of both a highly ordered magnetic field and a highly regular particle flow, it is possible for components of a PWN to have a shell-like morphology in this band (e.g., \cite{ng17}).

In addition, the synchrotron radiation from an individual particle will be linearly polarized.  Therefore, if the magnetic field inside a PWN is highly ordered -- as expected from theoretical work (e.g., Figures \ref{fig:pwn_mag} \& \ref{fig:stage1_turb}) -- the bulk radio emission will be significantly polarized as well.  Indeed, the radio emission from many PWNe is observed to be significantly polarized (linear polarization fractions $\gtrsim 5-30\%$; e.g., \cite{kothes06, kothes08, ng17, kothes20}) and the combination of central-filled radio morphology and significant linear polarization has been used to identify radio sources as PWN in the past (e.g. \cite{helfand84}).  In cases where polarized radio emission is not detected, this can indicate significant disorder, or anisotropies, in the PWN's magnetic field (e.g., Figures \ref{fig:pwn_mag} \& \ref{fig:stage1_turb}) and depolarization resulting from passage through material located near the PWN (e.g., \cite{weiler80}) or, more commonly, variations in the interstellar magnetic field along the line of sight (e.g., \cite{haverkorn04}) -- with the latter particularly relevant for more distant sources.

Additionally since the spectrum of the synchrotron radiation produced by an individual particle is strongly dependent on its energy (and the strength of the magnetic field at it's location, e.g., \cite{ginzburg65, ginzburg69, pacholczyk70, rybicki86}), the bulk radio spectrum observed from a PWN is indicative of the distribution in energy of the radiating leptons.  If the number of (radiating) particles $N$ with energy $E$ is given by a power-law:
\begin{eqnarray}
\label{eqn:epar-power}
\frac{dN}{dE} & \propto & E^{-p},
\end{eqnarray}
where $p$ is particle index, the flux density $S_\nu$ is expected to follow the relation
\begin{eqnarray}
\label{eqn:alpha}
S_\nu & \propto & \nu^\alpha,
\end{eqnarray}
where spectral index $\alpha$ is related to particle index $p$ by (e.g., \cite{ginzburg65, pacholczyk70, rybicki86}):
\begin{eqnarray}
\label{eqn:synch-relat}
\alpha & = & \frac{1-p}{2}.
\end{eqnarray}
Due to the low energies and long radiative lifetime of radio-emitting leptons, these particles are expected to be dominated by those previously injected into the PWN by the central neutron star, and therefore their spectrum reflects the evolution of the PWN during this time (e.g., \cite{hattori20}).  In general, the radio emission of PWN is characterized by a spectral index $\alpha \sim 0$ (see recent compilation by \cite{arad21})-- with this fairly unique combination of flat spectrum and center-filled morphology long used to identify such sources (e.g., \cite{becker83, helfand87}, even in cases where no pulsar has been detected (e.g., \cite{gelfand07}).  However, there are PWN with ``steep'' radio spectra, $\alpha \lesssim -0.5$, thought to arise from differences in the underlying particle acceleration mechanism, magnetic field strength, and/or evolutionary phase of the nebula (e.g., \cite{kothes08, kothes20}).

\subsection{Infrared, Optical, and Ultraviolet}
Similar to the radio emission described above, continuum emission in these wavebands generated within the PWN is again synchrotron radiation from relativistic leptons in this source. For younger PWN, where the radiative lifetime of these particles is larger than the age of these systems, this continuum emission often has a similar morphology to that observed in the radio (e.g., \cite{shklovsky58, weiler80, chanan84, woltjer87, blair97, shibanov08, sollerman00, slane08, dubner17}).  Furthermore, the continuum emission in these bands can often be significantly polarized (e.g., \cite{shklovsky58, zajczyk12}), as observed in the radio and again, consistent with the synchrotron radiation as the underlying emission mechanism.

As described above, the SED of the synchrotron radiation detected from a PWN is believed to be related to the energy distribution of the emitting particles.  While for some PWN, the flux density of their infrared to ultraviolet synchrotron emission lies along a power-law extrapolation of what is observed in the radio (e.g., \cite{slane08, hattori20}), a single power law often is not a good fit to the synchrotron emission observed in these wavebands and that in the radio (e.g., \cite{marsden84}), let alone within these bands themselves (e.g., \cite{lundqvist20, kim20}).  These spectral ``breaks" are often interpreted in the context of the energy distribution of particles injected into the PWN at the termination shock (e.g, Equation \ref{eqn:bpl}) or changes resulting from the evolution of particles within the nebula (e.g., radiative losses).  These evolutionary losses are also believed to responsible for the changes in spectral index $\alpha$ observed within PWN in these bands (e.g. \cite{vc93, temim06}).

Furthermore, the synchrotron radiation observed in these bands must be emitted from particles with higher energies (e.g., \cite{trumper70}) and/or in regions of stronger magnetic fields than their radio counterparts.  As such, the emission in these bands can often be dominated by structures near the termination shock, where particles are accelerated and then injected into the PWN.  Furthermore, hydrodynamic simulations (e.g., Figures \ref{fig:pwn_mag} \& \ref{fig:stage1_turb}) suggests this region is also where the nebular magnetic field is the strongest.  Therefore, the synchrotron radiation in these bands are often dominated by emission near this structure.  This is true for both older PWNe, where the super-sonic motion of the pulsar drives a bow shock into the surrounding medium as described above (e.g., Figure \ref{fig:bowshock}; \cite{wang13}), as well as younger PWN, where the compact structures associated with the termination shock are often apparent (e.g., \cite{shibanov08, slane08, zharikov13}) -- some of which are highly polarized (e.g., \cite{lundqvist11, moran13} and vary on timescales of days to weeks (e.g., \cite{tziamtzis09, lundqvist11}).

However, synchrotron radiation is not the only source of emission detected towards PWNe in these bands.  The infrared emission from many PWN is significantly above an extrapolation of the synchrotron radiation emitted in other wavebands (e.g., \cite{hattori20}).  For young PWN, this excess radiation is typically associated with dust formed inside the SNR and heated by, in most cases, the synchrotron emission from the PWN (e.g., \cite{marsden84, temim17, chawner19, priestley20, millard21}).  The physical properties (e.g., chemical composition, grain size, temperature, mass) of this material derived from its continuum and spectral line emission (e.g., \cite{temim06, chawner19, temim19, priestley20}) is important for understanding both the nature of the progenitor supernova and the formation of dust in these explosions.  This material also emits spectral lines observed at optical wavelength \cite{morse06, lundqvist22}, and the dense filaments formed as the PWN expands into the surrounding, cold supernova ejecta and dust (e.g., \cite{hester96}) can substantially absorb the optical and ultraviolet synchrotron radiative generated by the PWN within (e.g., \cite{sollerman00}).

\subsection{X-ray}
The X-ray emission detected from a PWN is again dominated by the synchrotron radiation from the high-energy leptons inside these sources -- though from higher energy particles, or more magnetized regions, than at lower photon energies.  Again, for PWNe where the synchrotron lifetime of these particles is longer than the true age of the system, and the extent and overall morphology of the X-ray emission is comparable to that observed at lower energies (e.g., \cite{hattori20}).

Again, the higher particle energies and stronger magnetic fields needed to generate synchrotron radiation at these photon energies results in this emission often being dominated by regions near the termination shock, which high angular resolution X-ray observations have revealed to have a complicated ring and jet-like structures (e.g., \cite{weisskopf00, pavlov03, weisskopf12}) in a wide variety of sources both young and old (e.g., \cite{lang10, pavlov06}).  The strong and highly ordered magnetic field associated with these features are expected to result in significantly polarized synchrotron emission, which has been detected from the Crab Nebula (e.g., \cite{weisskopf76, long21, hitomi18, bucciantini22}), with the detection of more being one of the goals of recent and upcoming missions (e.g., \cite{odell18, weisskopf22}).  The expected proximity between the X-ray emission of a PWN and the termination shock has been used to identify the location of the associated neutron star -- even in sources where it is not detected directly (e.g., \cite{gelfand07, slane12, temim13, temim15}), and an offset between the radio and X-ray emission of a PWN often taken as evidence for a collision between the PWN and reverse shock inside a SNR (e.g., \cite{temim15}). 
However, it is important to note that the observed X-ray morphology of a PWN depends strongly on capabilities of the observing instrument, primarily its sensitivity to large regions of low surface brightness -- with different observations of the same source yielding extremely different angular extents (e.g., \cite{lin12, matheson16}).  Therefore, extreme caution is needed when trying to estimate the physical properties of a PWN (e.g., its magnetic field strength) using differences in the observed size of the X-ray and radio emission of a PWN.

\subsection{Gamma-ray}

At gamma-ray energies above $\sim$ a few GeV, PWN are among the most common associations for Galactic sources. In many cases, however, it is unclear if the emission is generated by the PWN itself or by an associated SNR; these sources are typically designated `spp' or composite sources and can be more numerous than the sources designated `pwn' in gamma-ray catalogues to date \cite{2020ApJS..247...33A_4fgl}.  

This ambiguity arises due to the comparably poorer angular resolution of gamma-ray instruments (typically $\sim0.05^\circ-0.1^\circ$ above 1\,TeV at best) compared to radio and X-ray facilities (of order $\sim0.5-5''$ for the latter) \cite{2013APh....43....3A_CTAconcept}. The situation is further compounded by the different fields of view of X-ray facilities ($\sim30'$) and gamma-ray instruments ($\sim3^\circ-10^\circ$) \cite{2013APh....43....3A_CTAconcept}.
This challenge is inherent for gamma-ray astronomy due to the requirement for an increasing stopping power of the detector medium in the detection of higher energy photons. For increasing stopping power and rarity of very high energy photons, a larger collection area is required necessitating ground-based instruments rather than satellites. Consequently, as gamma-rays are detected via the particle cascades generated in the detector medium (e.g. the atmosphere or water at TeV energies), a fundamental limit to the angular resolution is imposed by the natural fluctuations in cascade development \cite{2006astro.ph..3076Hofmann}. Current instruments do not yet reach this limit, such that improvements are in principle possible, yet impose a challenge on efficient cascade detection and accurate timing \cite{2020APh...12302479Hofmann}. 

However, as is evident in table \ref{tab:halos_table}, gamma-ray emission associated to a specific PWN system often extends beyond the detected radio or X-ray PWN. A diffuse and low surface brightness component to the X-ray emission corresponding to the TeV extent has in general not been ruled out, yet the identification of such a component has been previously limited by the smaller field of view of X-ray instruments. The recently launched eROSITA satellite with a field-of-view of $1.03^\circ$ diameter may be able to address this question \cite{2021A&A...647A...1P_erosita}. 

\begin{table*}[h!]
\begin{center}
\begin{threeparttable}
\caption{Properties of selected well-known PWN systems in different evolutionary stages, ordered according to the system age. Table adapted from \cite{2020A&A...636A.113G}, see also references therein.  }
\begin{tabular}{l c c c c c c c c}
\hline \hline
System & Crab & MSH\,15-52 & G21.5-0.9 & G0.9+0.1 & Vela X & J1825-137  & Geminga\\
\hline      
Age (kyr)\tnote{a} & 0.94 & 1.56 & 4.85 & 5.31 & 11.3 & 21.4 & 342 \\
PSR\tnote{b} & {\small B0531+21} & {\small B1509-58} & {\small J1833-1034} & {\small J1747–2809} & {\small B0833-45} & {\small B1823-13} & {\small J0633+1746} \\ %
$\log(\dot E)$ (erg/s)& $38.65$ & $37.23$ & $37.53$ & $37.63$  & $36.84$ & $36.45$  & $34.51$ \\
Distance (kpc)& 2 & 4.4 & 4.1 & 8.5 & 0.28 & 3.93  & 0.25\\
R$_{\rm SNR}$ (pc)& ?\tnote{c} & 38.4 & 2.98 & 19.8 & 19.5 & 120  & ?\\
R$_{\rm PWN}$ (pc)\tnote{d} & 2.8 & 19.2 & 0.8 & 2.5 & 12.2 & ?  & 0.01\\
R$_{\rm TeV}$ (pc)\tnote{e} & $<$ 3 & 11 & $<$ 4 & $<$ 7 & 2.9  & 50  & 16.2\\
R$_{\rm X-ray}$ (pc)& 0.24 & 10.2 & 0.8 & 4.9 & 3.08 & 9.1  & 0.15\\
Stage\tnote{f} & 1 & 1 & 1b & 1b & 2 & 2b  & 3 \\
\hline
\end{tabular}
\begin{tablenotes}
\small
\item[a] The pulsar characteristic age is used for the age of the system, except where historical values are known. 
\item[b] Associated pulsar (PSR). Pulsar properties are taken from \cite{Manchester05}.
\item[c] Unknown quantities are marked by \lq\lq ?\rq\rq\/ 
\item[d] R$_{\rm PWN}$ is the size of the PWN in radio (as opposed to the radio SNR shell). 
\item[3] R$_{\rm TeV}$ is the one sigma radius taken from \cite{hess_pwn_pop} for sources within the H.E.S.S. Galactic Plane Survey (HGPS), unless a reference is provided. 
\item[f] Stage of system evolution is assigned loosely based on age, to correspond to Fig. \ref{fig:pwn_sketch}
\end{tablenotes}
\label{tab:halos_table}
\end{threeparttable}
\end{center}
\end{table*}

Broad-band emission from pulsar environments is typically comprised of synchrotron radiation and inverse Compton scattering spectral signatures. In the case of particularly energetic pulsars, the synchrotron self-Compton mechanism can also play an important role. 

\begin{figure}
    \centering
    \includegraphics[width=0.9\textwidth]{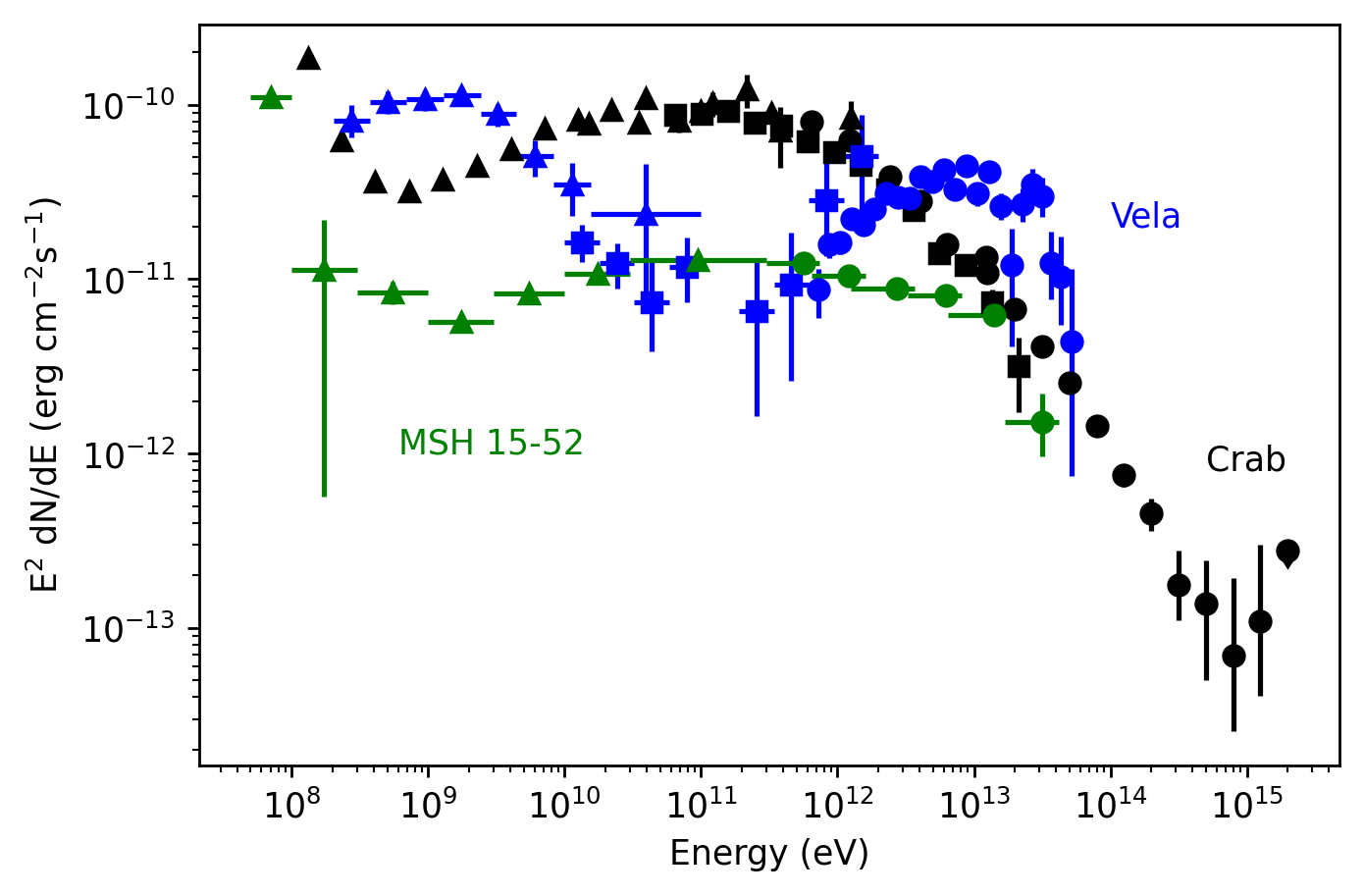}
    \caption{Spectral energy distributions from three pulsar wind nebulae; the Crab nebula \cite{2021Sci...373..425Lhaaso_crab,2020ApJ...897...33Arakawa,2015JHEAp...5...30A_magiccrab}, the Vela PWN \cite{2013ApJ...774..110G_Velax,velax_tibaldo,2019A&A...627A.100Hess_VelaX} and MSH\,15-52 \cite{2020ApJS..247...33A_4fgl,2018A&A...612A...1HGPS}. Measurements from the Fermi-LAT satellite are shown with triangles, whilst TeV measurements are shown with squares or circles.}
    \label{fig:pwn_sed}
\end{figure}

Spectral energy distributions (SEDs) from three PWNe in the range $\sim 0.1$\,GeV to $\sim1$\,PeV are shown in figure \ref{fig:pwn_sed}. Properties of these PWNe are summarised in table \ref{tab:halos_table}, adapted from \cite{2020A&A...636A.113G}. The Crab nebula is both the youngest and brightest PWN - as a result, it has been thoroughly studied at all wavelengths, leading to rich broad-band coverage of the emission. Most recently, LHAASO confirmed the detection of emission beyond 1\,PeV from the Crab nebula, indicating the presence of particles (leptons) with energies beyond the Cosmic Ray knee of $\sim3$\,PeV. 

MSH\,15-52 is a similarly young age, yet with lower flux normalisation of the inverse Compton component, likely due to a combination of factors; namely the distance to the PWN, the lower spin-down energy, and a different radiation field environment. By contrast, the Vela system clearly has an inverse Compton component with a peak shifted to much higher energies around 10\,TeV as opposed to $\sim$100\,GeV. 
As the distance to Vela is much lower, we are given a close-up view of the innermost regions of the PWN. The inverse Compton component is measured from a region known as Vela\,X, whereas emission from a broader region known as the extended radio nebula corresponds to an older electron population, which the highest-energy particles have already left via diffusive escape \cite{velax_escape}.

Despite the Vela pulsar having an older characteristic age than that of the Crab pulsar (PSR\,B0531+21) or PSR\,B1509-58 (powering the Crab nebula and MSH\,15-52 respectively), the peak of the inverse Compton emission occurs at higher energies due to the electron population within the measured region of Vela\,X being younger overall than the electron population powering the full nebula emission for the more distant Crab and MSH\,15-52 PWNe. As continuous accelerators, electrons (and positrons) released from the pulsar into the wind, nebula and eventually escaping into the surrounding medium are gradually transported via diffusion and/or advection, losing energy and cooling in the process. 

\begin{figure}
    \centering
    \includegraphics[width=0.8\textwidth]{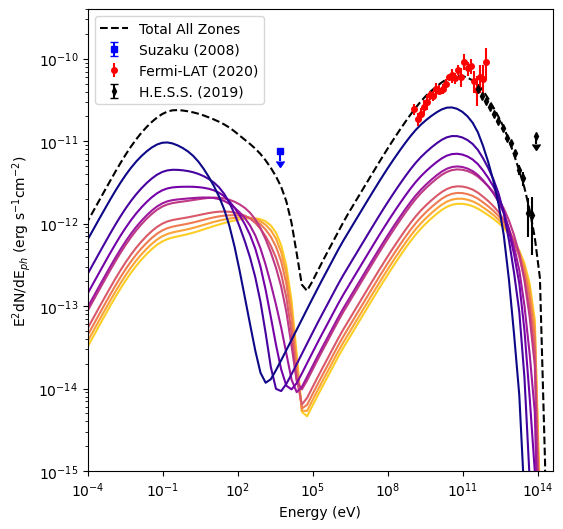}
    \caption{Spectral energy distribution showing how gamma-rays from an evolved pulsar wind nebula can be well described by a multi-zone model. The total measured emission is the sum of contributions from particle populations of different ages, with the youngest shown in yellow and oldest in blue. 
    Credit: Principe et al., A\&A, 640, A76, (2020) \cite{2020A&A...640A..76Principe}, reproduced with permission © ESO.
    }
    \label{fig:multizone}
\end{figure}

The effect of electron cooling with age since injection (release) is illustrated in figure \ref{fig:multizone}, adapted from \cite{2020A&A...640A..76Principe}. 
As electrons cool from the youngest (yellow) to the oldest (blue) zones, the peak of the emission shifts to lower energy. Synchrotron emission decreases the peak energy due to decreasing magnetic field with increasing distance. Similarly, particle energy losses cause the inverse Compton peak to shift towards lower energies, whilst the energy loss per scattering interaction decreases leading to an increased total gamma-ray flux at lower energies. Summing together the contributions from all zones provides a reasonable description of the total measured emission \citep{2011ApJ...742...62V,2020A&A...640A..76Principe}. 

\begin{figure}
    \centering
    \includegraphics[width=0.49\textwidth]{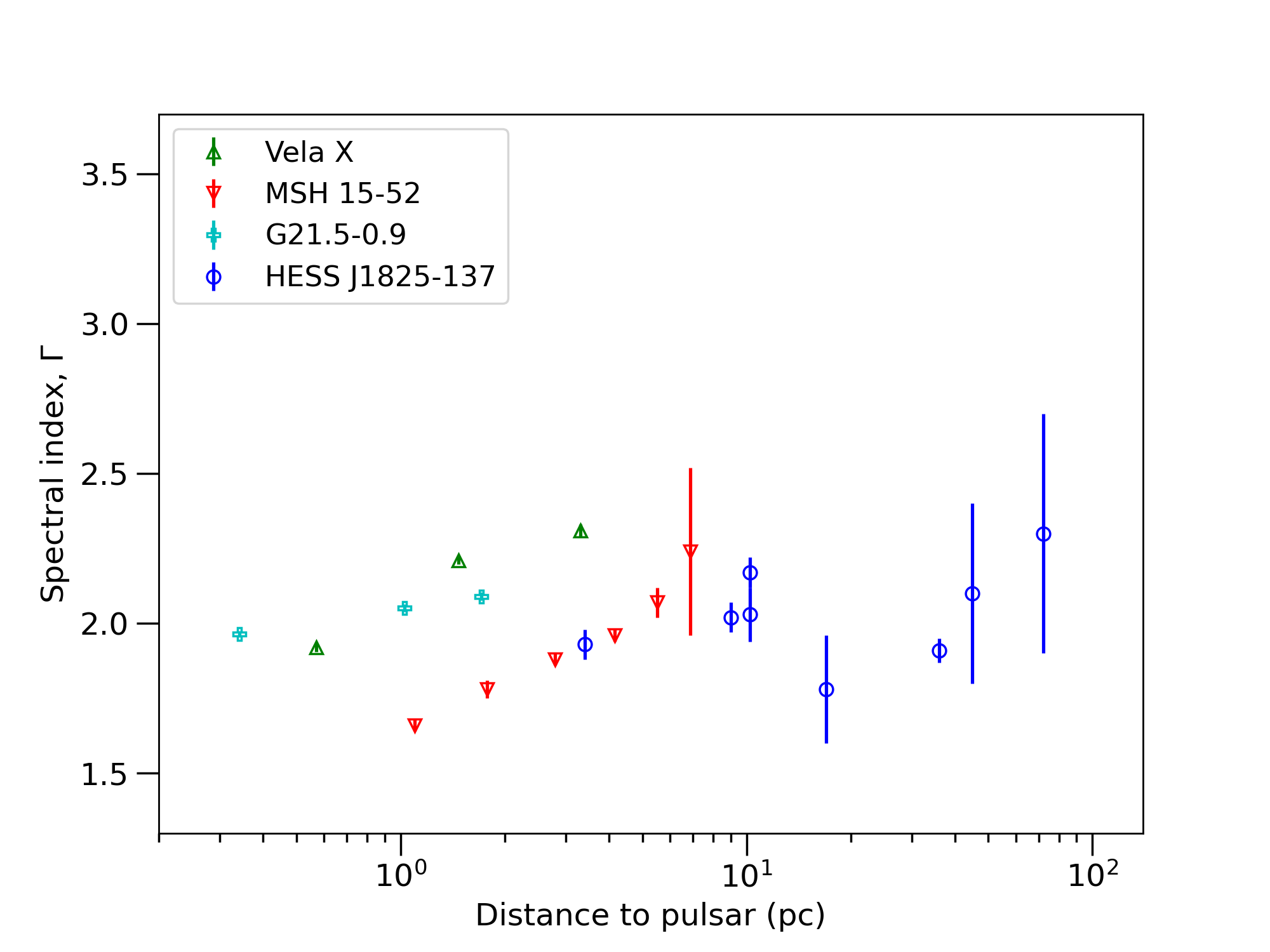}
    \includegraphics[width=0.49\textwidth]{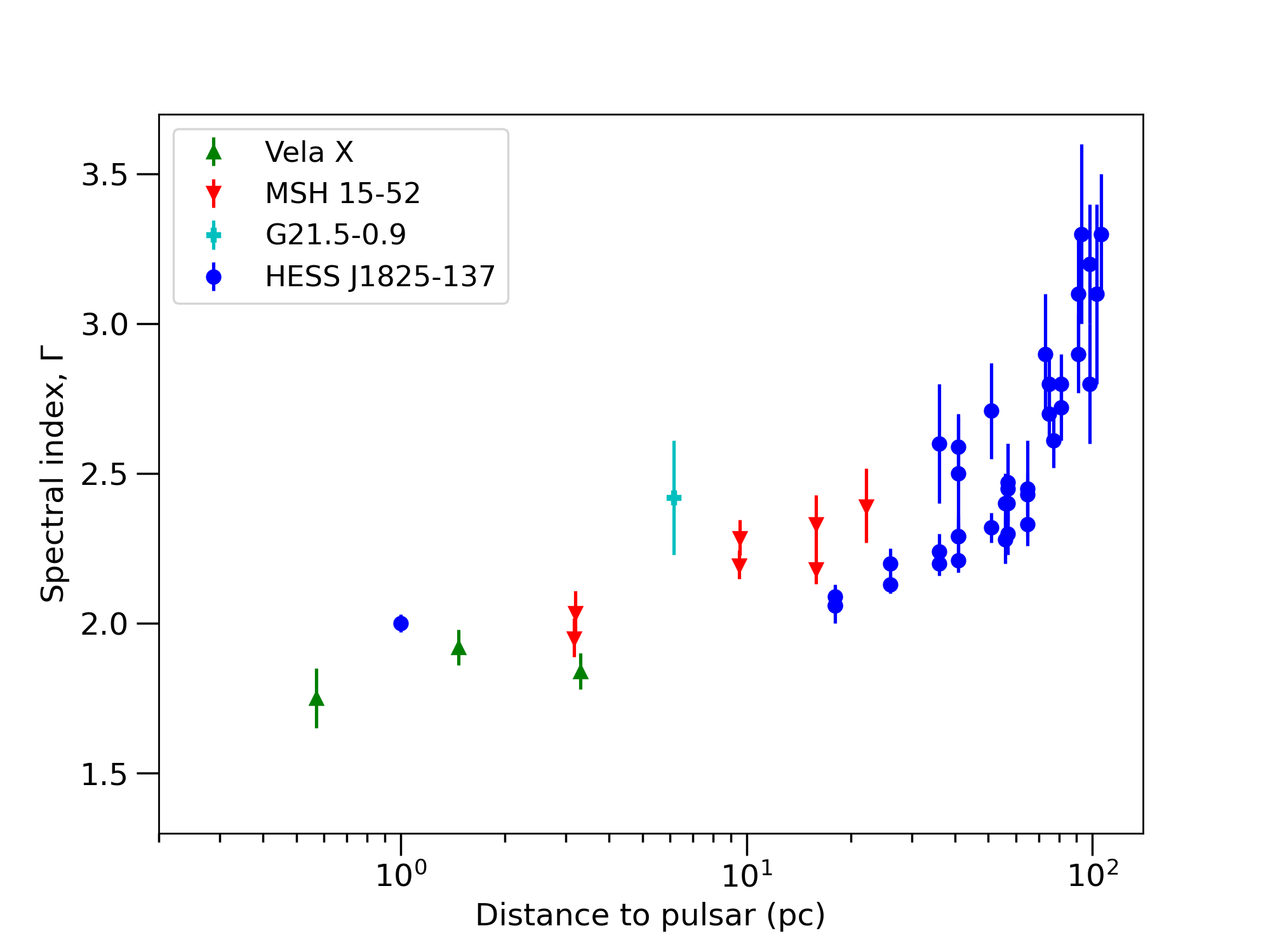}
    \caption{ Power law spectral index with distance from pulsar, shown for (left) X-ray and (right) TeV gamma-ray emission. Data compiled in both X-ray and gamma-ray for four canonical PWNe: Vela\,X \citep{2019A&A...627A.100Hess_VelaX}, MSH\,15-52 \citep{2010A&A...515A.109Schock,Hillert}, G21.5-0.9 \citep{2014ApJ...789...72NustarG21,2018A&A...612A...1HGPS} and HESS\,J1825-137 \citep{2011ApJ...742...62V,2019A&A...621A.116H} }
    \label{fig:PLind_radius}
\end{figure}

Figure \ref{fig:PLind_radius} shows the variation in spectral index with distance from the pulsar for both Synchrotron emission in the X-ray band and inverse Compton emission in the TeV band. 
A changing spectral index for a power law fit with increasing physical distance from the pulsar illustrates nicely the effect of particle cooling with time and transport through the nebula and surrounding medium, corroborating the model shown in figure \ref{fig:multizone}. The general trend for PWNe in both X-ray and gamma-ray domains is for spectral index that softens with increasing distance from the pulsar. 
Significant energy-dependent morphology within the gamma-ray regime is hence a signature for PWNe. 

\subsection{Young PWN: the crab nebula}

\begin{figure}
    \centering
    \includegraphics[width=0.48\textwidth,trim=5cm 7cm 5cm 7cm,clip]{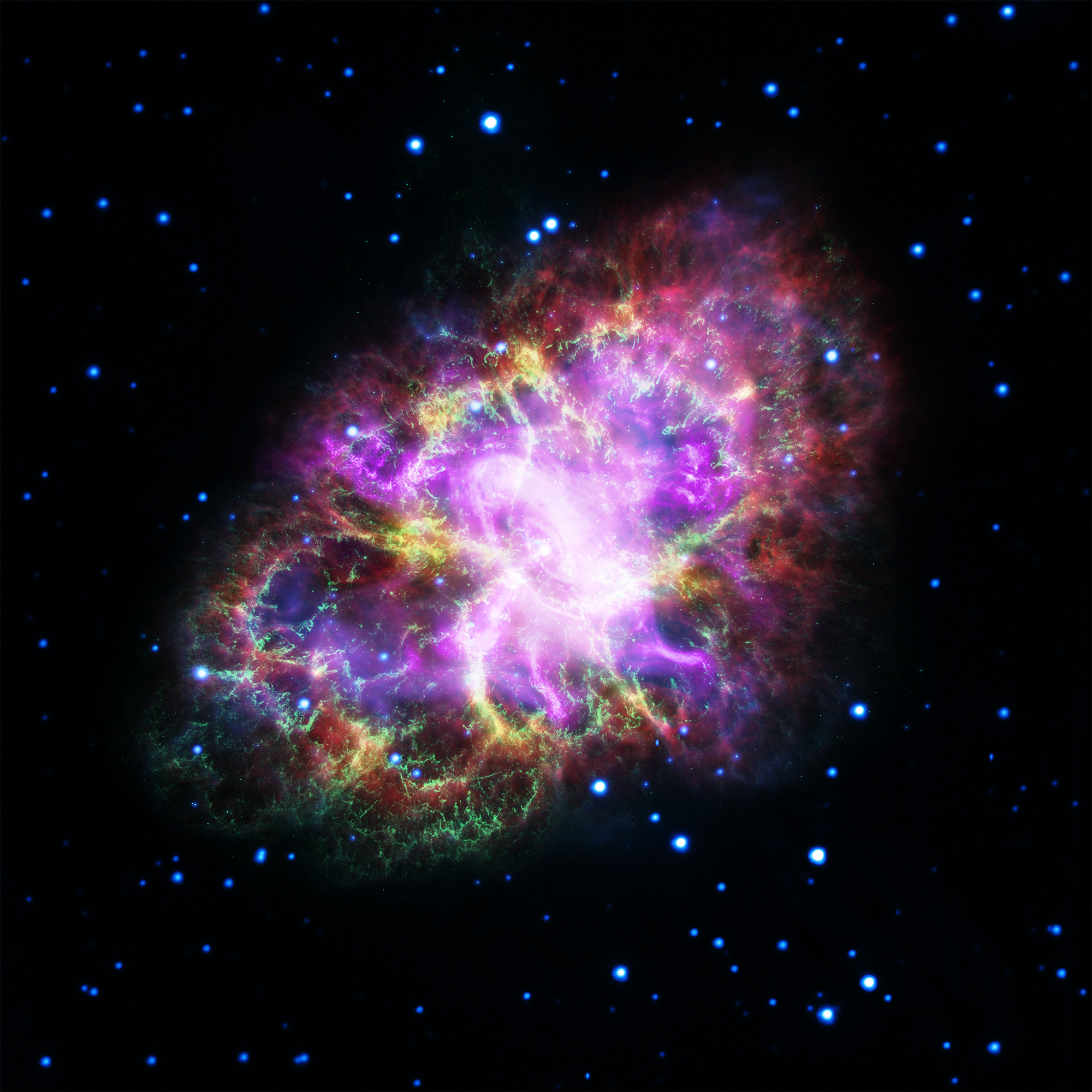}
    \includegraphics[width=0.48\textwidth]{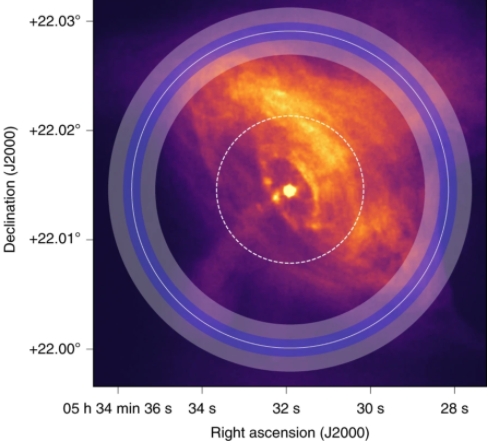}
    \caption{Left: Composite image showing the Crab nebula as seen in optical with the X-ray jets and toroidal structure shown in lilac. Right: Close up of the X-ray emission from the Crab pulsar, with the TeV gamma-ray extent measured by the H.E.S.S. array shown (with error bands) by a white circle and hard X-ray extent measured by Chandra shown by a dashed circle \cite{2020NatAs...4..167HessCrab}. }
    \label{fig:crab}
\end{figure}

In 1054 a supernova explosion was observed and recorded by Chinese astronomers. The stellar remnant formed in the collapse is the Crab pulsar, which continues to power the bright and well-known Crab Nebula. A composite image is shown in figure \ref{fig:crab}, including optical and X-ray (purple) emission. The zoom-in of the X-ray torus also indicates the extent of the TeV gamma-ray emission, overlaid by a ring including uncertainties and centred on the best-fit centroid of the emission \cite{2020NatAs...4..167HessCrab}. The 52 arcsecond radius extent of the TeV emission is much larger than the extent of hard X-ray measured by Chandra, indicated by a white dashed circle. Nevertheless, this marks one of the most precise resolutions yet achieved at TeV energies. Variation in the measured extent with energy is accounted for by energy-dependent radiation losses; the most energetic electrons are found close to the pulsar, whilst electrons lose energy as they propagate outwards, such that less energetic electrons can be found out to larger radii. Hence, the TeV inverse Compton emission occupies a region with larger radius than that of the hard X-ray synchrotron emission, yet a smaller radius than that of the UV synchrotron \cite{2020NatAs...4..167HessCrab}.

\subsection{`Stage 2': Vela X}

As outlined above, the system around the Vela pulsar is especially interesting on account of its close proximity to Earth, providing us with a close-up view of the inner nebula and various components of the pulsar environment. In particular, the spectral maximum of the inverse Compton peak occurs at an energy among the highest known for all TeV gamma-ray sources. 

\begin{figure}
    \centering
    \includegraphics[width=\textwidth]{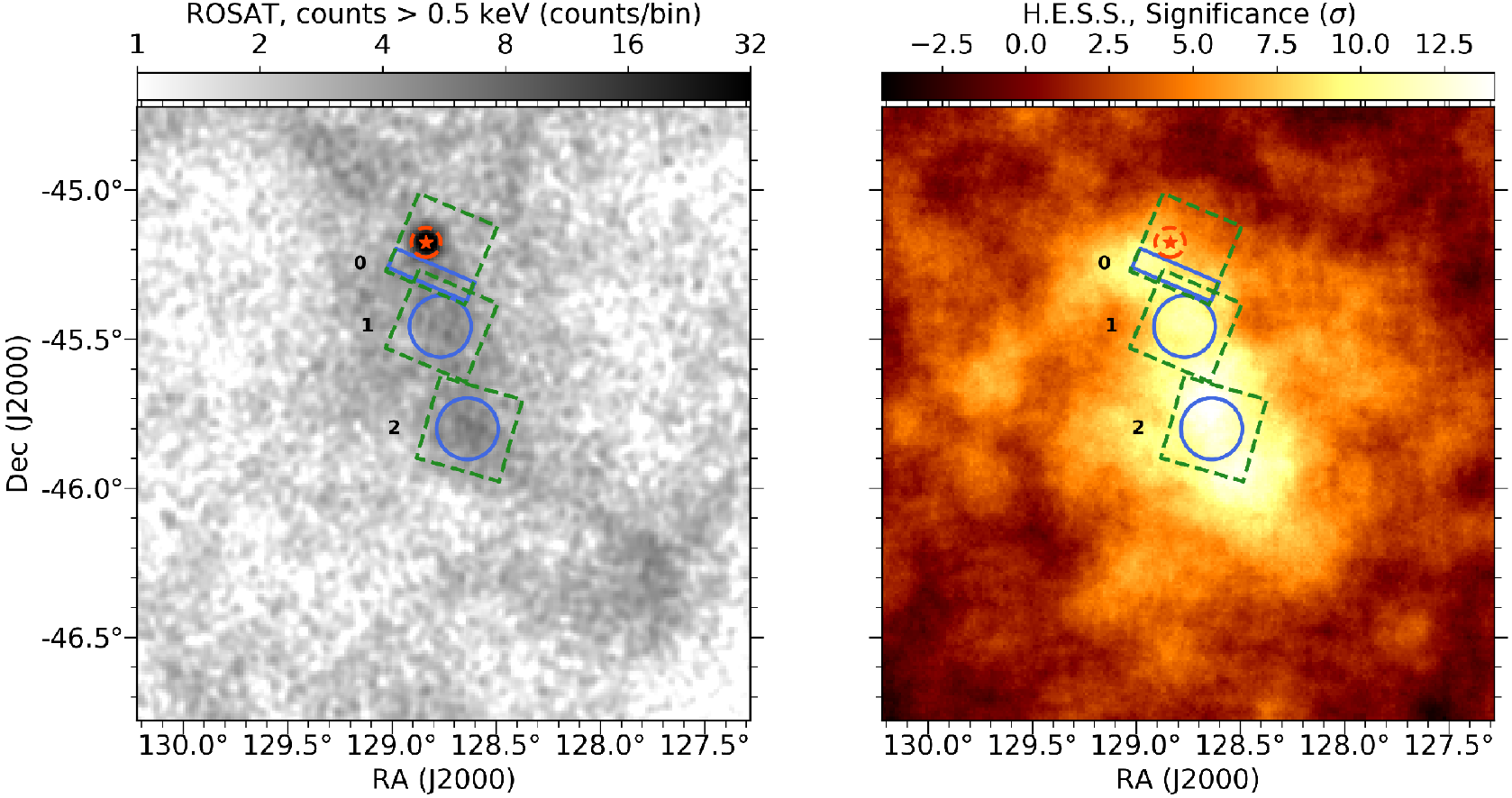}
    \caption{X-ray and gamma-ray images of the Vela\,X PWN, with the location of the Vela pulsar indicated by a red star in both images.
    Credit: H.E.S.S. Collaboration, A\&A, 627, A100, (2019), reproduced with permission © ESO.
    }
    \label{fig:velax}
\end{figure}

Spectra extracted from the locations indicated in figure \ref{fig:velax} were used to constrain the magnetic field strength within the nebula \cite{2019A&A...627A.100Hess_VelaX}, with the corresponding spectral indices plotted in figure \ref{fig:PLind_radius}. 
Typical Stage 2 systems are characterised by their transitional state, with some particle escape underway, yet a considerable concentration of energetic particles remain close to the pulsar \cite{velax_escape}. 

Further examples of stage 2 PWNe include CTA\,1 and G292.2-0.5, both powered by young pulsars with spin-down energy of $\dot{E}=4.5\times10^{35}$erg\,s$^{-1}$ (PSR\,J0007+7303, $13$\,kyr) and $\dot{E}=2.3\times10^{36}$erg\,s$^{-1}$ (PSR\,J1119-6127, $1.6$\,kyr) respectively \cite{Manchester05}. For these two systems, associated TeV emission is seen extending beyond the size of the PWN in X-ray and interactions between the SNR reverse and PWN forward shock are thought to have occurred, affecting the resulting morphology \cite{2013ApJ...764...38A_CTA1,2018A&A...612A...1HGPS}. PSR\,J1119-6127 is notable for both its comparatively high magnetic field and for having a measured braking index of $n=2.684\pm0.0002$ which would increase the pulsar age slightly, although this remains younger than the age of the parent SNR. Due to the limited angular resolution of gamma-ray instruments, it remains ambiguous whether the TeV emission associated with the composite system, HESS\,J1119-614, originates from the SNR or from the PWN \cite{2018A&A...612A...1HGPS}.

\subsubsection{Pulsar Halos}

Recently, halos of relativistic particles (electrons and positrons) diffusing into the ISM have been identified as a new type of source through their signature gamma-ray emission produced through inverse Compton scattering (see \cite{2022NatAs...6..199L} for a recent review). 
Initially, pulsar halos were proposed as a phenomenon distinct from the canonical PWN in the evolution of pulsar environments, characterised by their diffusive escape from the PWN \citep{2017PhRvD..96j3016L}. However, such a distinction also requires apriori knowledge of (or assumptions about) the state of the system; whether particles are trapped with the PWN (leading to a centrally peaked surface brightness profile) or the system has already become disrupted such that particles are free to escape into the surrounding medium. 

A complementary approach is to estimate the energy density in energetic particles that are present in the region \citep{2020A&A...636A.113G}. For regions in which the energy density of energetic particles is considerably higher than the ISM average, the accelerator can be considered to dominate the dynamics of the surrounding medium (corresponding to pulsar wind nebulae). Where the energy density of energetic particles is considerably lower than the ISM average, the accelerator no longer dominates the dynamics of the surrounding medium. This applies to halos of escaped particles - although the presence of highly energetic particles can still be inferred, the energy density is no longer sufficient to be dominant over the energy density of other ISM components. As such, the escaped particles are dispersed into the surrounding medium, forming a `halo'. 

An additional complication with such a classification is that many such systems can be observed in a transitional state; where the system has become partially disrupted, but the majority of particles are not yet travelling freely in the ISM, or the energetics change with distance from the pulsar. 
Therefore, the precise classification of a specific source may be considered less important than our general understanding of the evolution of such pulsar environments. 

The phenomenon of such `halos' - of particles escaped from the accelerating region - is not unique to pulsars but can be replicated around other accelerators, such as supernova remnants or stellar clusters \cite{2021A&A...654A.139Brose}. 

\subsection{`Middle-aged': Geminga}

\begin{figure}
    \centering
    \includegraphics[width=0.8\textwidth]{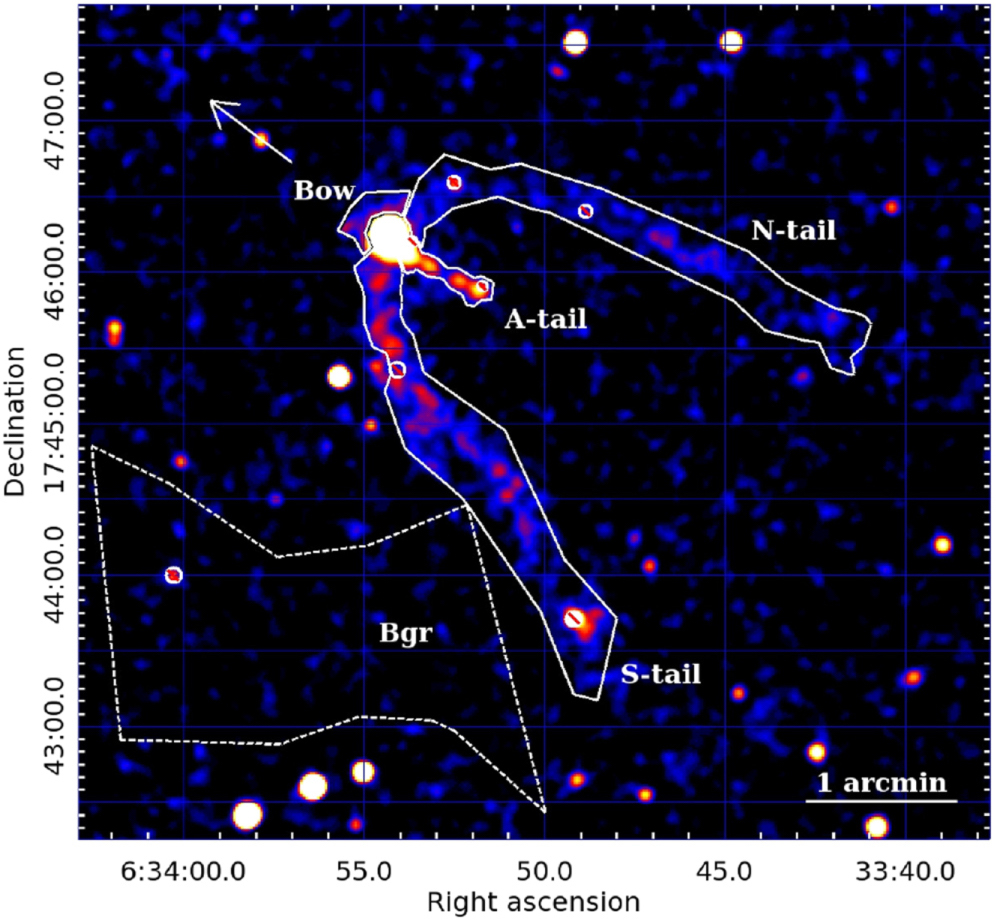}
    \caption{Geminga pulsar and tails seen in X-ray; as a stage 3 system, the pulsar is escaped in the ISM, with tails trailing in the opposite direction to the pulsar proper motion. Credit: Posselt et al., ApJ, 835, 66 (2017) \cite{geminga_posselt_xray}, © AAS. Reproduced by permission. } 
    \label{fig:geminga}
\end{figure}

Geminga - as the archetypal `pulsar halo' system - has one of the most extreme size ratios between the X-ray and gamma-ray extents (see table \ref{tab:halos_table}). Whilst the TeV gamma-ray emission extends over a broad region several tens of pc in radius, the X-ray emission is confined to below 1\,pc in extent, with highly ordered structure in the form of tails following the magnetic field lines (figure \ref{fig:geminga}). From the angular scale, we see that the length of the X-ray tails is approximately 10 arc seconds - in contrast, the TeV emission extends to at least 5 degrees radius \cite{HAWCgeminga_2017Sci...358..911A}. 
This considerable discrepancy is due to the age of the electrons producing the corresponding electromagnetic emission. Younger, more recently accelerated electrons and positrons are trapped along the strong magnetic field lines, producing synchrotron X-ray emission. 
Older, relic electrons continue to scatter off ambient radiation fields producing inverse Compton gamma-ray emission over a much longer period of time, gradually dispersing into the ISM. An additional effect - relevant for comparatively nearby pulsars such as Geminga, located at a mere $\sim 250$\,pc - is the proper motion of the pulsar over time. Particles that were released earlier during the pulsars history may continue to be centred around the former location of the pulsar long after the pulsars trajectory has led to a `new' apparent location on the sky \cite{diMauroFermi2019PhRvD.100l3015D,2020A&A...640A..76Principe}. 

\begin{figure}
    \centering
    \includegraphics[width=0.9\textwidth]{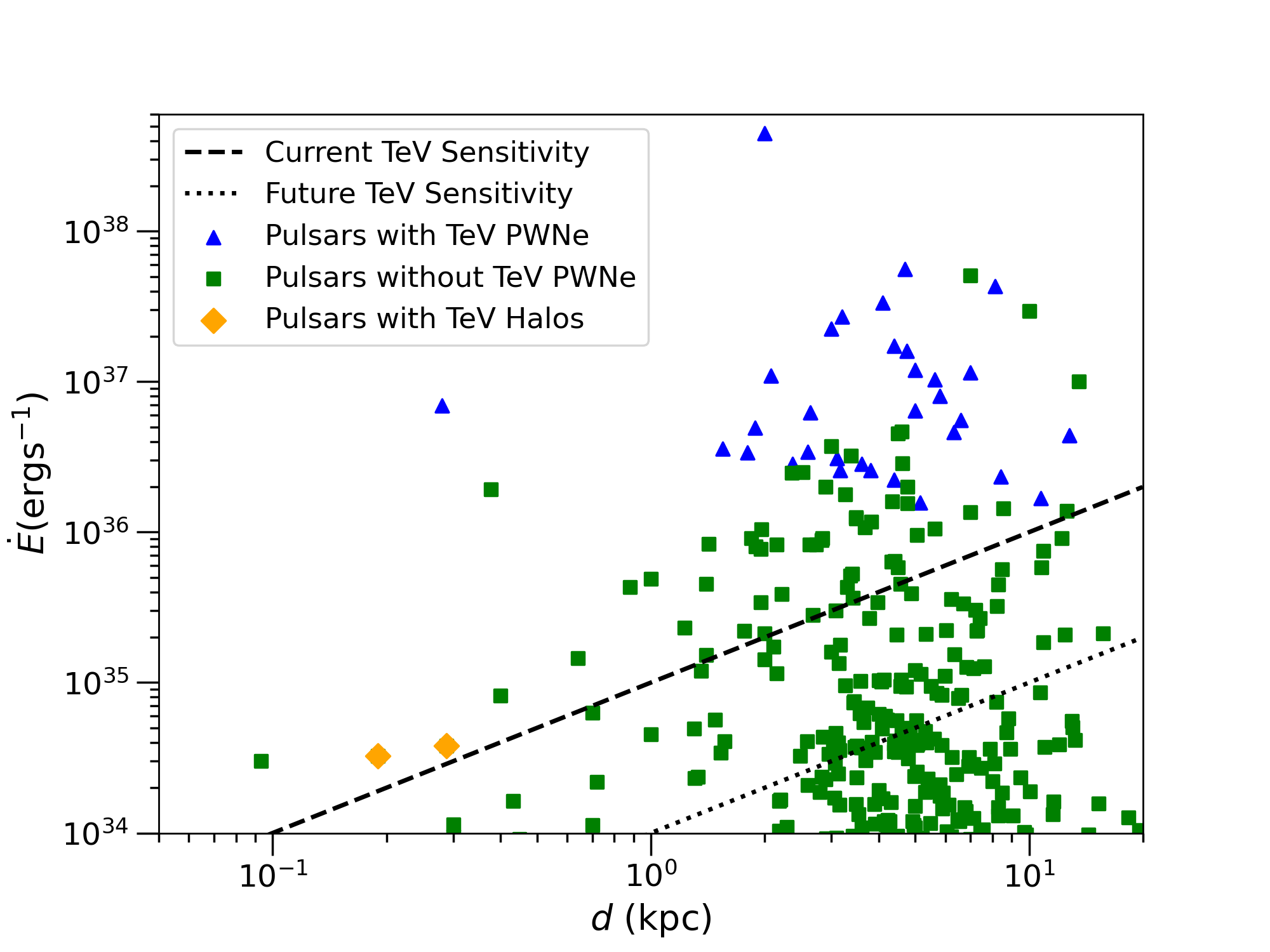}
    \caption{Energetic pulsar population showing those for which TeV associations have been established and the projected sensitivity of future instruments.}
    \label{fig:tev_pop}
\end{figure}

So far, only two systems are generally accepted as true pulsar halos; namely the extended TeV emission around the Geminga pulsar and around PSR\,B0656+14, although several other VHE sources have been suggested as candidate halos \cite{2017ATel10941....1R,2018ATel12013....1B,2021PhRvL.126x1103A_halo}. Nevertheless, there are many pulsars known whose properties (such as spin-down power, proximity to Earth) suggest that they may be harbouring an as yet undetected PWN or halo. Figure \ref{fig:tev_pop} provides an overview of the known energetic pulsar population from \cite{Manchester05}, indicating which pulsars have known TeV detected PWN or halo counterparts. It is also clear that the anticipated improvement in sensitivity from currently operational to future instruments will potentially unlock a large number of systems for detection, particularly among the Stage 3 of development, when a large halo of escaped particles could form.  

Although only two pulsars have confirmed halos detected at TeV energies, there is an ongoing investigation as to how many of the canonical PWNe may deserve a reclassification as either halo systems comprised of part PWN, part halo. 
Providing a classification based on measured properties of the emission must distinguish between the intrinsic properties of the PWN and halo components - despite the fact that these will most commonly overlap along the line of sight (see figure \ref{fig:pwn_sketch}). 



\subsection{Ultra-High-Energy gamma-ray emission} 

\begin{figure}
    \centering
    \includegraphics[width=0.9\textwidth]{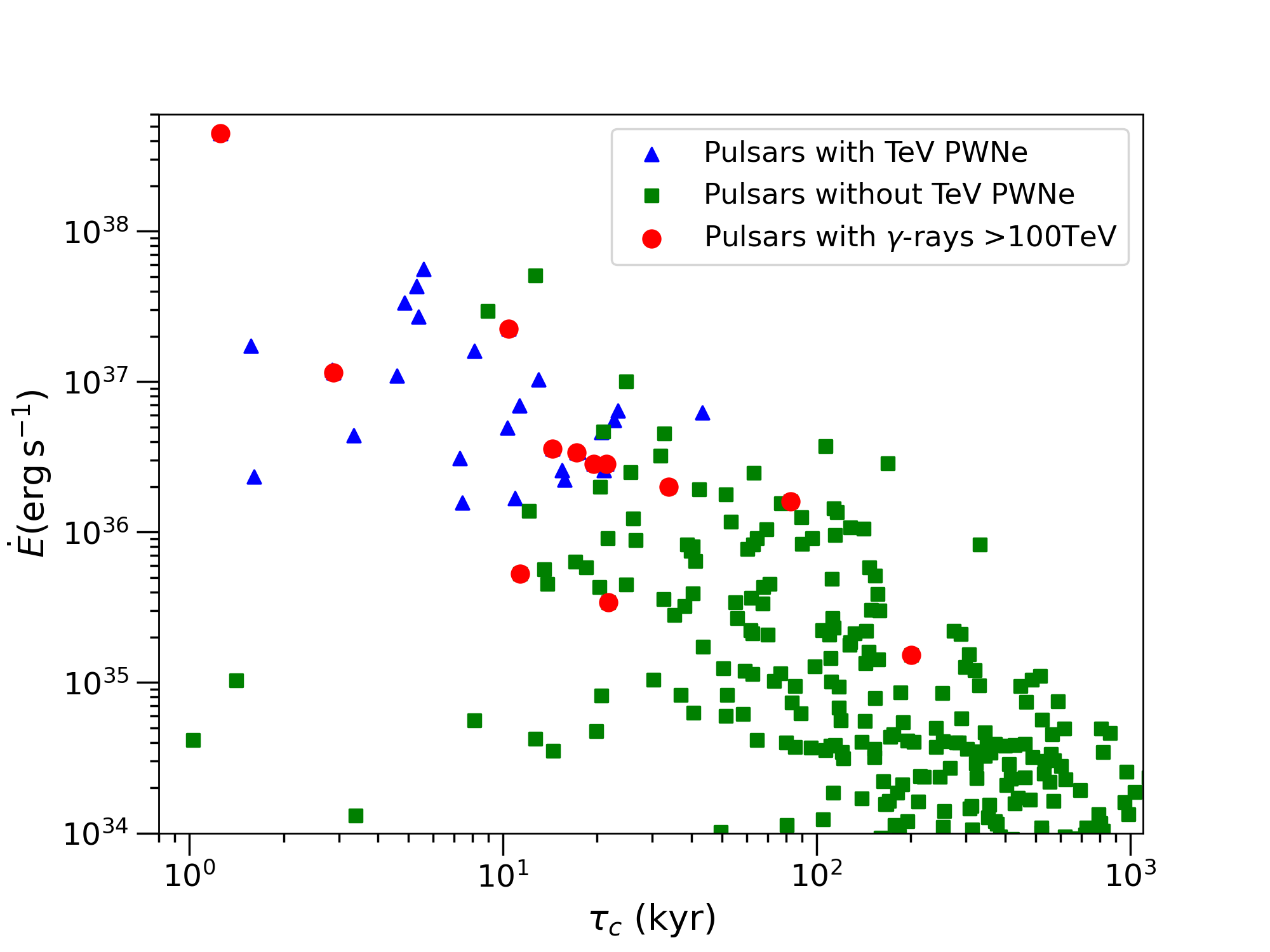}
    \caption{ Energetic pulsar population highlighting those from which ``ultra-high-energy'' emission beyond 100\,TeV has been measured. }
    \label{fig:uhe_pwn}
\end{figure}

Recent results from the Water Cherenkov Detectors HAWC and LHAASO have revealed a population of gamma-ray sources that continue to emit above $\sim100$\,TeV, christened `ultra-high-energy' (UHE) sources \cite{PhysRevLett.124.021102_hawc56,2021Natur.594...33C}. Most of these sources have plausible associations with known pulsars, and are highlighted in figure \ref{fig:uhe_pwn}. One of the first points to note is that - apart from the Crab with its spin-down power $\dot{E} = 4.5\times10^{38}\mathrm{
erg\,s}^{-1}$ - these pulsars are not exceptional with respect to the pulsar population. This implies that either there is something special about these pulsars and their surrounding PWN configuration that we have not yet identified, or that particle acceleration to UHE (above $\sim1$\,PeV, corresponding to gamma-rays $\sim100$\,TeV) is more common-place than previously thought. It is worth also noting, however, that in some cases the association with a nearby pulsar is tenuous and a different cause may be responsible for the observed UHE emission. 

The majority of the gamma-ray emission from PWNe is expected to be generated through Inverse Compton (IC) scattering. At energies beyond a few tens of TeV, the Klein-Nishina effect become relevant; this is the scattering regime in which electrons lose a significant fraction of their energy on each scattering interaction, such that the IC spectrum cuts-off rapidly. Emission at energies beyond this cut-off could hint at the presence of an underlying hadronic particle population, which is less affected by this cut-off \cite{1998nspt.conf..439A_ppcrab}. However, it has recently been shown that the Klein-Nishina effect is also mitigated for electrons in high radiation field environments \cite{2021ApJ...908L..49Breuhaus}. As yet, there is therefore no compelling reason to invoke a hadronic particle population to account for the highest energy emission observed in PWNe. 

\section{ Recent progress  and open questions}

\subsection{PWNe as PeVatrons}
PWNe are established as leptonic PeVatrons - accelerators of electrons and positrons to PeV energies; however whether or not they can also be considered cosmic ray PeVatrons - accelerators of hadronic particles to PeV energies - remains an open question. 
From the Hillas criterion, we can obtain an approximate estimate of the maximum energy achievable in a given astrophysical environment to first order. For Neutron Stars, energies well in excess of the cosmic ray knee at $\sim1$\,PeV and up to $10^{21}$eV are within reach by this simple approach, due to their strong magnetic fields. 

The next question to address is how hadronic particles could be present in PWNe, for which there are two plausible mechanisms. 
Firstly, ions extracted from the Neutron Star surface itself. 
These charged particles co-rotate with the pulsar at small radii, producing synchrotron radiation in the strong magnetic fields. The synchrotron photons are sufficiently energetic to undergo pair-production, the chain of synchrotron radiation and pair-production leading to a pair production  multiplicity factor in the range $10^2$ to $10^4$ \cite{2001ApJ...560..871H_arons}. Leptonic pairs hence dwarf the ions in number by the pulsar wind termination shock, which marks the start of the nebula. 
Although the nebula is subsequently dominated by pairs, it remains feasible that a small fraction of ions also survive to the pulsar wind termination shock - even undergoing further acceleration there - in order to enter the nebula \cite{2020A&A...635A.138Guepin,2015JCAP...08..026Kotera,lemoine15}. 

A second approach by which hadronic particles may be accelerated in a PWN is through shock mixing. As the parent SNR develops, a reverse shock forms that returns inwards towards the PWN and can re-introduce hadronic particles to the PWN forward shock. It has previously been shown that via this mechanism, energy gains of a factor $\sim$few up to $\sim100$ are possible, with maximum energies reaching beyond a PeV \cite{1992MNRAS.257..493B,2018MNRAS.478..926O}.

\subsection{`Non-Pulsar' Wind Nebulae}
As mentioned earlier, PWNe are believed to be powered by the rotational energy of an associated neutron star.  While the most common manifestation of these objects are rotation-powered pulsars, where the pulsed emission is also powered by the rotational energy, there exist neutron stars whose spin-down luminosity is less than their neutron bolometric luminosity (e.g., Anomalous X-ray Pulsars, Isolated Neutron Stars) or insufficient to explain to the transient behavior exhibited by these objects (e.g., Soft $\gamma$ Repeaters).  However, even for these neutron stars, the loss of rotational energy is expected to produce a pulsar wind similar to that observed from rotation-powered pulsars.  Extended X-ray emission around some "magnetars" (e.g., \cite{vink09, younes12, younes16}) -- isolated neutron stars whose emission is believed to driven by the evolution of extremely strong ($\gtrsim10^{14}~{\rm G}$) surface magnetic fields -- have been associated with PWNe (e.g, \cite{torres17}), i.e. powered by their rotational energy, though other origins of the emitting plasma are possible (e.g. \cite{hao16, granot17}).  In addition, some neutron stars associated with PWNe (e.g., \cite{ng08, kumar12}) have shown evidence for magnetar-like behavior (e.g., \cite{gavriil08, blumer21}), which is believed to be responsible for observed changes in the PWN's properties (e.g., \cite{blumer17, reynolds18}).

\subsection{Particle transport (diffusion and advection)}
\label{sec:transport}


Much research has been dedicated in recent years to the study of particle transport both within and beyond pulsar wind nebulae. Internal to a PWN, particle transport may provide via advective or diffusive processes, or most commonly a combination thereof. In a study of the energy dependent radius of HESS\,J1825-137, an extended gamma-ray source powered by the pulsar PSR\,B1823-13, the HESS collaboration found indications that advection is the preferred scenario for the bulk / dominant particle motion along the central axis, although this does not exclude an additional diffusive component to the motion \cite{2019A&A...621A.116H}.

Although slow diffusion within PWNe is well established, once particles have escaped into the surrounding ISM (such as within a pulsar halo), it was expected that the transport is subsequently governed by the properties of the ISM. It was therefore a surprise to find that the diffusion even within this halo of escaped particles is slow with respect to the Galactic average as inferred from the cosmic ray B/C ratio measurements \cite{2007ARNPS..57..285Strong,HAWCgeminga_2017Sci...358..911A}. 
If such a low diffusion coefficient does accurately reflect the average value over the intervening ISM, then the concept of pulsars as the most likely source of the positron excess and highest energy electrons is challenged. 

\begin{figure}
    \centering
    \includegraphics[width=0.7\textwidth]{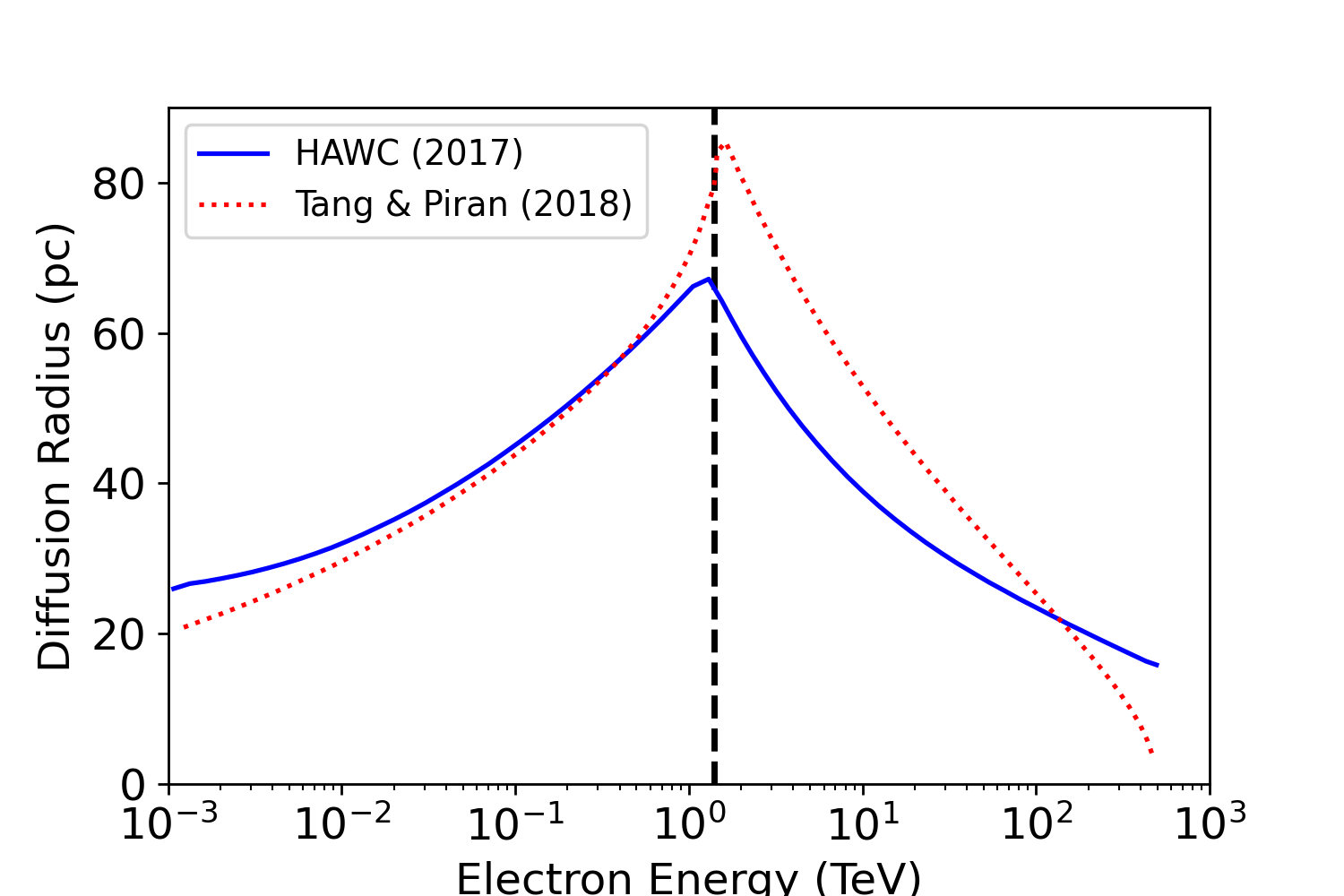}
    \caption{ Diffusion radius of electrons as a function of energy for two different transport models \cite{HAWCgeminga_2017Sci...358..911A,Tang2019MNRAS.484.3491T} both assuming a diffusion coefficient of $D=4.5\times 10^{27}\left(\frac{E}{\mathrm{100\,TeV}}\right)^{1/3} \mathrm{cm}^2/\mathrm{s}$, yet with different energy loss descriptions and approximations to the energy dependence. }
    \label{fig:Diff_Radius}
\end{figure}

Assuming that particles propagate away from the pulsar diffusively and symmetrically, the observed radial surface brightness profile can be fit with a diffusion model to obtain the best fit diffusion coefficient $D$. In \cite{HAWCgeminga_2017Sci...358..911A}, this was done for the extended emission around the Geminga pulsar obtaining a best fit value of $D$ that is low compared with the established Galactic average ISM value. 

Figure \ref{fig:Diff_Radius} describes the evolution of the diffusion radius with electron energy in the case of the Geminga pulsar; despite using the same $D_0$, visible differences between the two model curves are due to differences in treatment of the solution to the diffusion equation - solved numerically in \cite{Tang2019MNRAS.484.3491T} but with an analytical approximation used in \cite{HAWCgeminga_2017Sci...358..911A}.
The peak, located at $\sim 1.5$\,TeV, is defined by the energy at which the cooling time for electrons equals the age of the pulsar; namely 342\,kyr for Geminga, and is hence the same for both curves.  

Consequently, many studies have sought to explain this slow diffusion effect and the origin of the positron excess under scenarios including: i) a slow diffusion zone within the halo caused by turbulence due to the particles themselves, with the diffusion coefficient increasing to Galactic average values outside this zone \cite{Evoli18}; ii) the presence of an as yet undiscovered pulsar in the local vicinity \cite{LopezUnPulsar2018PhRvL.121y1106L}; iii) a combination of quasi-ballistic flow and diffusive effects \cite{2021PhRvD.104l3017Recchia}; among other explanations.  

Additionally, many studies have investigated whether the presence of so-called `Slow Diffusion Zones' (SDZ) around pulsars are in fact ubiquitous \cite{2020PhRvD.101j3035D_evidences} although so far such studies have fallen short of circumventing the intrinsic difficulty of distinguishing observationally between the canonical PWN and escaped particle halo components. SDZs may have implications not only for the local positron excess and high energy electron flux, but also for Cosmic Ray propagation throughout the entire Galaxy \cite{Johannesson2019ApJ...879...91J}.

\section{Conclusion}

Although many years of studies and detailed observations have led to significant progress in our understanding of pulsar wind nebulae, several open questions remain. 
(i) What is the pair-production multiplicity factor and what proportion of the pulsar wind is carried by ions? (ii) Is the observed discrepancy between the angular size of a PWN between different wavebands (in particular, X-rays and $\gamma$-rays) an observational artefact or of physical significance (e.g., magnetic field structure, spatial distribution as a function of particle energy)?  (iii) How are particles transported through the nebula and does the particle diffusion appear to be inhibited in the halo region around the pulsar?

New facilities both in X-rays and gamma-rays promise to help answer these questions; in particular, the current eROSITA telescope with its large field-of-view will significantly improve the sensitivity to extended and diffuse X-ray emission \cite{2021A&A...647A...1P_erosita}. Additionally, first results from IXPE have started to demonstrate the power of X-ray polarimetry for studies of PWNe, shedding important information on the geometry of the magnetic field at the termination shock \cite{bucciantini22}.
In the gamma-ray regime, future ground-based facilities are foreseen using both aforementioned techniques of Water Cherenkov Detectors such as the Southern Wide-field Gamma-ray Observatory (SWGO) and imaging atmospheric Cherenkov telescopes such as the Cherenkov Telescope Array (CTA) Observatory. The former will continually survey the sky and is well suited to identifying regions of significant and highly extended emission - as expected for stage 3 systems harbouring large halos of energetic particles. Whereas the latter, CTA, will have good angular and energy resolution capabilities and is hence highly suited to detailed spectro-morphological studies of specific PWN systems. 

Combining multi-wavelength information for a PWN together with physical models for the evolution of these sources will ultimately provide the most complete insights into the processes at work.  This is required for understanding the mechanisms responsible for some of the highest energy events and particles observed in the our Galaxy. 

\newpage

\section*{Acknowledgements}

\noindent AM is supported by the Deutsche Forschungsgemeinschaft (DFG, German Research Foundation) – Project Number 452934793.




\bibliographystyle{ieeetr}
\bibliography{PWNchapter}

\end{document}